\documentclass[reprint,prd,nofootinbib,superscriptaddress,twocolumn]{revtex4}
\usepackage{amsfonts}
\usepackage{amsmath}
\usepackage{amssymb}
\usepackage[english]{babel}
\usepackage{graphicx}
\usepackage{epsfig}
\usepackage{bm}
\usepackage{verbatim}
\usepackage[utf8]{inputenc}
\usepackage{booktabs}
\usepackage{multirow}
\usepackage{subfigure}
\usepackage{xcolor}
\usepackage[colorlinks=true,urlcolor=red,citecolor=red]{hyperref}
\usepackage[font=small]{caption}
\usepackage{float}
\usepackage{blindtext}
\usepackage{booktabs}
\usepackage{mathrsfs}

\newcommand{\bc}{\begin{center}}
\newcommand{\ec}{\end{center}}

\newcommand{\mathsym}[1]{}

\newcommand{\paren}[1]{\left(#1\right)}

\begin{document}

\title{
{Relation between pole and running masses of heavy quarks using the principle of maximum conformality}\\
}
\author{Daniel Salinas-Arizmendi}
\email{daniel.salinas@usm.cl}
\author{Iván Schmidt}
\email{ivan.schmidt@usm.cl}
\affiliation{Departamento de Física, Universidad T\'{e}cnica Federico Santa Mar\'{\i}a, y Centro Científico-Tecnológico de Valparaíso,
Casilla 110-V, Valpara\'{\i}so, Chile}
\date{\today }

\begin{abstract}
The relation of the pole and running heavy quark masses of order $\mathcal{O}\left(\alpha_s^4\right)$ in perturbative quantum chromodynamics (pQCD) can be obtained using the Principle of Maximum Conformality (PMC), a formalism that provides a rigorous method for eliminating renormalization scale and scheme ambiguities for observables in pQCD. Using PMC, an optimal renormalization scale for the heavy quark mass ratio is determined, independent of the renormalization scale and scheme up to order $\alpha_s^4$.
Precise values are then obtained for the PMC pole masses of the heavy quarks $M_b^{\text{PMC}}=4.86^{+0.03}_{-0.02}$ GeV, $M_t^{\text{PMC}}=172.3\pm 0.6$ GeV, and the running mass $\overline{m}_t^{\text{PMC}}=162.6\pm 0.7$ GeV at the PMC scale.
\end{abstract}
\maketitle

\section{\label{intro}Introduction}
The mass of the quarks plays an important role in the phenomenology of high energy physics. For example, the $b$ quark mass is essential for determining the decay of the $B$ meson and is the dominant channel in the decay of the Higgs boson into a pair of quarks, while the top quark is used indirectly to determine the mass of the Higgs boson.

In lowest order of perturbation theory, there is no need to choose a renormalization scheme when considering quantum corrections to  quark masses, which are defined differently depending on the renormalization scheme. One of the most widely used renormalization schemes is the modified minimal subtraction scheme, $\overline{\text{MS}}$, which corresponds to a specific choice of the finite parts of the counterterms. Then the renormalized mass and coupling constant explicitly depend on the renormalization scale, through the renormalization group equations (RGE), i.e. of the order of the scale of the process. Another widely used renormalization scheme is the on-shell subtraction scheme, where the renormalized mass is defined in all orders as the position of the pole in the particle propagator.

High-order QCD corrections to the quark mass ratio have been calculated up to $\mathcal{O}\paren{\alpha_s^3}$ \cite{gray1990three,melnikov2000three} and $\mathcal{O}\paren{\alpha_s^ 4}$ \cite{marquard2015quark, Steinhauser2016xkf,kataev2020multiloop}, who argue that the renormalization scale should be set to the typical momentum scale of the process to eliminate large logarithms; this estimated scale is then varied over an arbitrary range to determine its uncertainty. However, this conventional procedure gives scheme-dependent predictions and thus violates the fundamental principle of renormalization group invariance. 
%
%
The pQCD series for the relation between the pole mass and the mass in other schemes has a problem of convergence, as the contributions of higher orders  grow factorially, as $n! \paren{\beta_0 \alpha_s}^n$.
This problem is not always solved within the conventional renormalization procedure, where the choice of subtraction scheme and scale are not free from ambiguities. 

The Principle of Maximum Conformality (PMC) \cite{brodsky2012scale,brodsky2012setting, brodsky2012eliminating,mojaza2013systematic, brodsky2014systematic,wu2013renormalization,singlescale1}, which underlies the Brodsky-Lepage-Mackenzie (BLM) method \cite{brodsky1983elimination},
represents the best alternative to solve the problem exposed in the previous paragraph, by providing a systematic way to remove renormalization scheme and scale ambiguities. The PMC method determines the renormalization scale by absorbing all the non-conformal terms governing the running coupling behavior, and thus the resulting pQCD series is conformal.

\section{Formalism}

The starting point of the PMC method is the introduction of the $\delta$-$\mathcal{R}$enomalization scheme \cite{mojaza2013systematic,brodsky2014systematic},
 which is a generalization of the conventional scheme used in dimensional regularization \cite{bollini1972dimensional,veltman1972regularization}, where a constant $-\delta$ is subtracted in addition to the standard $\ln 4\pi - \gamma_E$ subtraction of the $\overline{\text{MS}}$ scheme:
 \begin{equation}
\mu^2=\mu_{\delta}^2 \exp\left( \ln(4\pi)-\gamma_E-\delta\right).
\end{equation}
The subtraction $\delta$ defines an infinite set of renormalization schemes, where the physical results are independent of $\delta$, i.e. scheme independent. Since the PMC predictions do not depend on the choice of renormalization scheme, the PMC scaling satisfies the RGE invariance principles \cite{brodsky2012self}.

The PMC procedure \cite{brodsky2014systematic}  consists first of choosing a renormalization scheme, $\mathcal{R}_\delta$ for example, 
and a renormalization scale. 
The invariance of the RGE ensures the connection between different schemes by the evolution of the coupling constant in terms of the $\beta$-function, through the renormalization group equation:

\begin{equation} \label{RGE1}
 \frac{d }{d \ln \mu_\delta^2} \left[ \frac{\alpha_s^{\mathcal{R}}\left(\mu_\delta\right)}{\pi} \right]=\beta\left(\frac{\alpha^{\mathcal{R}}_s\left(\mu_\delta\right)}{\pi}\right),
\end{equation}
where the beta function is
\begin{equation}
\beta\left(\frac{\alpha^{\mathcal{R}}_s\left(\mu_\delta\right)}{\pi}\right)=-\sum_{n\geq 0} \beta_n  \times \left(\frac{\alpha^{\mathcal{R}}_s\left(\mu_\delta\right)}{\pi}\right)^{n+2},
\end{equation}

The first two $\beta_n$ terms do not depend on the choice of the renormalization scheme.
For higher orders, to simplify the notation, we introduce $a_s = \alpha_s^{\mathcal{R}}/\pi$. An approximate analytical solution is obtained by integrating Eq.\eqref{RGE1}


\begin{equation}
\begin{aligned}
\ell_{\delta}
= & \int^{a_s(\mu_\delta)} \frac{d a}{\beta\left(a(\mu)\right)}=\frac{1}{\beta_0}\left[\frac{1}{a_s}+b_1 \ln a_s \right.\\
& +a_s\left(-b_1^2+b_2\right) +a_s^2\left(\frac{b_1^3}{2}-b_1 b_2+\frac{b_3}{2}\right) \\
&  +a_s^3\left(-\frac{b_1^4}{3} 
 +b_1^2 b_2-\frac{b_2^2}{3}-\frac{2}{3} b_1 b_3  +\frac{b_4}{3}\right)\\
 &  +\mathcal{O}\left(a_s^4\right)\bigg]+B,
\end{aligned}
\end{equation}

where $\ell_{\delta} = \ln \mu_{\delta}^2/\Lambda_{\text{QCD}}^2$, 
$b_i=\beta_i/\beta_0$ with $i=1,\dots,4$, and $B$ an arbitrary integration constant. Through an iterative inversion, at the next-to-next-to-leading-order (NNLO), the result for the coupling is obtained:
%
%
\begin{equation}
\begin{aligned}
a_s  = & \frac{1}{ \ell_{\delta} } - b_1 \frac{\ln\ell_{\delta}}{\ell_{\delta}^2} + \frac{1}{\ell_{\delta}^3} \left[b_1^2 \left(\ln^2 \ell_{\delta} - \ln \ell_{\delta}-1\right) +b_2\right] \\
 & + \frac{1}{\ell_{\delta}^4} \left[ b_1^3 \left( -\ln^3 \ell_{\delta} + \frac{5}{2} \ln ^2 \ell_{\delta} + 2 \ln \ell_{\delta}- \frac{1}{2}\right) \right.\\ 
 & \left. -3b_1b_2 \ln \ell_{\delta} +\frac{b_3}{2} \right]
 + \frac{1}{\ell_{\delta}^5} \left[b_1^4 \left( \ln^4  \ell_{\delta} -\frac{13}{3} \ln^3\ell_{\delta} \right. \right.\\
&  \left. -\frac{3}{2} \ln^2 \ell_{\delta} + 4 \ln \ell_{\delta}+ \frac{7}{6}\right) +3b_1^2b_2  \left(2\ln^2\ell_{\delta} \right. \\
& \left. \left. -\ln \ell_{\delta}-1 \right)-b_1 b_3 \left( 2 \ln \ell_{\delta} + \frac{1}{6} \right) + \frac{5}{3} b_2^2 +\frac{b_4}{3}  \right].
\end{aligned}
\end{equation}

An observable in pQCD in any renormalization scheme can be written as 

\begin{equation}
\mathscr{P}_0(\mu_\delta) =t_0 +\sum_{i\geq 0} t_{i+1}\left(\mu_\delta ;\mu_0 \right) a_s^{i+p} (\mu_0),
\label{observable}
\end{equation}

where $\mu_0$ is an initial renormalization scale and $p$ is a loop index representing the power of the coupling associated with the tree-level term. In the ideal case, the observable is independent of the renormalization scale, which is achieved by fully summing the series. In practice, however, this is not feasible due to the mathematical complexity of the higher-order terms. By truncating the sum, the series becomes sensitive to the renormalization scale and to a particular scheme, introducing a first ambiguity conditioned by theoretical conventions. Using the coupling displacement relation, in any \hbox{$\mathcal{R}$-scheme} a result given in two different scales can be connected to each other 
\begin{equation} \label{relationalpha}
a_s\paren{\mu}=a_s\paren{\mu_\delta} + \sum_{k\geq 1}
\frac{1}{k!} \left.\frac{d^k a_s\paren{\mu}}{\paren{d \ln \mu^2}^k}\right\vert_{\mu=\mu_\delta} \paren{\ln\frac{\mu^2}{\mu_\delta^2}}^k.
\end{equation}

Taking into account the displacement relation, the 
observable of Eq.\eqref{observable} can be written as
\begin{equation} \label{Obs1}
\begin{aligned}
\mathscr{P}_{0} =  & \ t_0 +t_1a_s^p +a_{s}^{p+1}\left( t_{2}+p\beta_{0}t_{1}
\ln \frac{\mu_{\delta}}{\mu_0}\right)  
 \\
&+  a_s^{p+2} \bigg( t_{3} \bigg.  +p\beta _{1}t_{1}\ln \frac{\mu_{\delta}}{\mu_0}+\left( p+1\right) \beta
_{0}t_{2}\ln \frac{\mu_{\delta}}{\mu_0}\\
& \left. +\frac{p\left( p+1\right) }{2}\beta _{0}^{2}t_{1}\ln^2\frac{\mu_\delta}{\mu_0}\right) +a_s^{p+3} \bigg( t_{4} \bigg. \\
&  +p\beta _{2}t_{1}\ln \frac{\mu_{\delta}}{\mu_0} +\left( p+1\right) \beta
_{1}t_{2}\ln \frac{\mu_{\delta}}{\mu_0} \\
&+\left( p+2\right) \beta _{0}t_{3}\ln \frac{\mu_{\delta}}{\mu_0} +\frac{p\left( 3+2p\right) }{2}\beta _{0}\beta _{1}t_{1}\ln^2 \frac{\mu_{\delta}}{\mu_0}  \\
& + \frac{\left( p+1\right) \left( p+2\right) }{2}\beta_{0}^{2}t_{2}\ln^2 \frac{\mu_{\delta}}{\mu_0}\\
&\left. +\frac{p\left( p+1\right) \left( p+2\right) }{3!}\beta _{0}^{3}t_{1}\ln^3 \frac{\mu_{\delta}}{\mu_0}\right) + \mathcal{O}(a_s^{p+4}).
\end{aligned}
\end{equation}

The RGE invariance is essential for obtaining consistent physical predictions, and its fulfillment requires adjustments of the scales in the running coupling to eliminate nonconformal terms associated with the $\beta$ function. In a conformal theory (${\beta}= 0$), scale dependence disappears, so by absorbing all $\beta$ dependence into the effective coupling at every order, one can obtain a final result independent of the initial choice of scale and scheme (up to that order). The use of a $\mathcal{R}$ scheme \cite{brodsky2014systematic} provides a rigorous foundation for this process. This approach ensures that the remaining conformal terms are scheme independent, and the numerical validity of the prediction at finite orders is scheme independent, in accordance with renormalization group principles. This scheme invariance criterion, which is crucial at any truncated order of the perturbative series, differs from the formal statement that the all-order expression for a physical observable is scale and scheme invariant. The final result obtained represents the conformal theory, proved by PMC \cite{brodsky2014systematic}, which prescribes the resummation of all nonconformal terms in the perturbative series in the running coupling.

Eq.\eqref{Obs1} reveals a pattern in the coefficients of the terms at each order. By replacing the coefficients $t_i$ with $t_{i,j}$, the observable is written as

\begin{equation} \label{Obs2}
\begin{aligned}
\mathscr{P} _0 =& \ t_0 + t_{1,0}a_s^{p}(Q)+\left[
t_{2,0}+p\beta _{0}t_{2,1}\right] a_s^{p+1}(Q) \\
& +\bigg[ t_{3,0}+p\beta _{1}t_{2,1}+\left( p+1\right) \beta _{0}t_{3,1} \bigg. \\
& \left.  +\frac{p\left( p+1\right) }{2}\beta _{0}^{2}t_{3,2}\right] a_s^{p+2}(Q) + \bigg[t_{4,0}  +p\beta _{2}t_{2,1}\bigg.\\
& +\left( p+1\right) \beta_{1}t_{3,1}+\left(p+2\right) \beta _{0}t_{4,1}\\
& +\frac{p\left( 3+2p\right) }{2}\beta _{0}\beta _{1}t_{3,2} + \frac{\left( p+1\right) \left( p+2\right) }{2}\beta_{0}^{2}t_{2,1} \\
&\left.+ \frac{p\left( p+1\right) \left( p+2\right) }{3!}\beta
_{0}^{3}t_{4,3}\right] a_s^{p+3}(Q) + \mathcal{O}(a_s^{p+4}).
\end{aligned}
\end{equation}

In this notation, the conformal and nonconformal parts of the perturbative coefficients are denoted by $t_{i,0}$ and $t_{i,j}$, respectively.

In the expression of Eq.\eqref{Obs2}, all linear terms in ${\beta}$ can be summed up in $t_{i,1}$ by defining new scales $Q_i$ at each order. This can be expressed as

\begin{equation} \label{redef1}
\begin{aligned}
t_{1,0}a_s\left( Q_{1}\right) ^{p} =&t_{1,0}a_s\left( Q\right) ^{p}-p\ a_s\left(
Q\right) ^{p-1}\beta \left( a_s\right) t_{2,1} \\
t_{2,0}a_s\left( Q_{2}\right) ^{p+1} =&t_{2,0}a_s\left( Q\right) ^{p+1}-\left(
p+1\right) a_s\left( Q\right) ^{p}\beta \left( a_s\right) t_{3,1} \\
t_{3,0}a_s\left( Q_{3}\right) ^{p+2} =&t_{3,0}a_s\left( Q\right) ^{p+2}-\left(
p+2\right) a_s\left( Q\right) ^{p+1}\beta \left( a_s\right) t_{4,1} \\
&\vdots  \\
t_{k,0}a_s\left( Q_{k}\right) ^{k} =&t_{k,0}a_s\left( Q\right) ^{k}-ka\left(
Q\right) ^{k-1}\beta \left( a_s\right) t_{k+1,1}
\end{aligned}
\end{equation}

At the same time, from the displacement relation given in Eq. \eqref{relationalpha}, the running coupling is expressed in terms of $a_s\left( Q_{k}\right) ^{k}$:
\begin{widetext}
\begin{equation} \label{redef2}
a_s\left( Q_{k}\right) ^{k} = \ a_s\left( Q\right) ^{k}+k\ a_s\left( Q\right)
^{k-1}\beta \left( a_s\right) \ln \frac{Q_{k}^{2}}{Q^{2}} 
+\frac{k}{2}a_s\left( Q\right) ^{k-2}\left[ \beta \frac{d\beta }{da_s}a_s\left(
Q\right)  \right. 
\left. +\left( k-1\right) \beta \left( a_s\right) ^{2}\right] \ln ^{2}\frac{Q_{k}^{2}}{Q^{2}}+\cdots
\end{equation}
\end{widetext}

Considering the $k^{\text{th}}$ power of the running coupling from Eq.\eqref{redef1} and Eq.\eqref{redef2}, it is found that the new renormalization scale satisfies

\begin{equation}
-\frac{t_{k+1,1}}{t_{k, 0}}=\ln \frac{Q_k^2}{Q^2}+\frac{1}{2}\left[\frac{\partial \beta}{\partial a_s}+(k-1) \frac{\beta}{a_s}\right] \ln ^2 \frac{Q_k^2}{Q^2}+\cdots
\end{equation}

Therefore, the PMC procedure simultaneously determines the 
scale at each order \cite{mojaza2013systematic}. 
The PMC scales $Q_k$ are determined up to N$^3$LO order by the following formula (detailed derivations are given elsewhere \cite{brodsky2014systematic}):

\begin{widetext}
\begin{equation}
\ln \left(\frac{ Q_{k}}{Q} \right)^2=\frac{%
\tau_{k,1}+\Delta _{k}^{\left( 1\right) }\left(
a_{s}\right) \tau_{k,2}+\Delta _{k}^{\left( 2\right)
}\left( a_{s}\right) \tau_{k,3}}{1+\Delta _{k}^{\left(
1\right) }\left( a_{s}\right) \tau_{k,1}+\left( \Delta
_{k}^{\left( 1\right) }\left( a_{s}\right) \right) ^{2}\left(
\tau_{k,2}-\tau_{k,1}\right) +\Delta
_{k}^{\left( 2\right) }\left( a_{s}\right)  \tau_{k,1}^{2}},  \label{multiscale}
\end{equation}
\end{widetext}
where 
\begin{eqnarray}
\tau_{k,j}&=&\left( -1\right) ^{j}\frac{%
t_{k+j,j}}{t_{k,0}}, \label{rrr}\\
\Delta _{k}^{\left( 1\right) }\left( a_{s}\right)  &=&\frac{1}{2!}\left[ 
\frac{\partial \beta }{\partial a_{s}}+\left( k+p-2\right) \frac{\beta }{%
a_{s}}\right]. \label{rrr2} \\
\Delta _{k}^{\left( 2\right) }\left( a_{s}\right)  &=&\frac{1}{3!}\left[
\beta \frac{\partial ^{2}\beta }{\partial a_{s}^{2}}+\left( \frac{\partial
\beta }{\partial a_{s}}\right) ^{2}\right. \label{rrr3}\\
& & +3\left( k+p-2\right) \frac{\beta }{a_{s}}%
\frac{\partial \beta }{\partial a_{s}}   \notag \\
\nonumber &&\left. +\left( k+p-2\right) \left( k+p-3\right) \frac{\beta ^{2}}{a_{s}^{2}%
}\right],
\end{eqnarray}%

The final pQCD prediction for $\mathscr{P}$ after setting the PMC
scales $Q_k$ then reads:

\begin{equation}
\begin{aligned}
\mathscr{P}^{\text{PMC}} =& \ t_{0,0} + t_{1,0} a_s(Q_1)^p +  t_{2,0} a_s(Q_2)^{p+1} + t_{3,0} a_s(Q_3)^{p+2}\\
 & + t_{4,0} a_s(Q_4)^{p+3} + \mathcal{O}(a_s^{p+4}).
\end{aligned}
\end{equation}

PMC scales are perturbative in nature, so we need to know more loop terms to get more accurate predictions, so $Q_4$ is unknown since we need the terms $\{\beta_i\}$ of order $a_s^{p+4}$. However, $Q_4$ can be set as the last determined scale $Q_3$, which guarantees the scheme independence of our prediction 
according to Ref.
\cite{CSR}.

\section{Quark mass relations in perturbative QCD}


In the present work, the relation between the pole mass and the running mass for heavy quarks is studied using the PMC method. The obtained PMC mass is independent of the choice of the initial subtraction scheme chosen for the calculations, up to the order of the expansion. 

The main pole- and running-mass relation \cite{Steinhauser2016xkf,marquard2016ms,kataev2020multiloop,Kataev:2016jai,Kataev:2015gvt} results from the renormalization of the bare mass $m_0$, which is given by
\begin{equation} 
m_0=Z^{\text{OS}}\ M,\quad \text{and} \quad m_0=Z^{\overline{\text{MS}}}\ \overline{m},
\end{equation}
where $M$ and $\overline{m}$ are the pole and $\overline{\text{MS}}$ quark masses, respectively. Next, we consider the relation between the pole mass and the running mass of the heavy quarks, namely
\begin{equation}\label{m0}
H_{q}\left( \mu \right) =\frac{M }{\overline{m}\left( \mu \right) }
= \frac{Z^{\overline{\text{MS}}}}{Z^{\text{OS}}}.
\end{equation}

The renormalized expression in Eq.\eqref{m0} can be expressed through a standard QCD perturbative series as
%

\begin{equation} \label{cm1}
H_{q}\left( \mu \right) =\frac{M }{\overline{m} \left( \mu \right)}%
=1+\sum_{n\geq 1}h_{n} \left[a_{s}(\mu)\right] ^{n},
\end{equation}

where is the order $n$ of perturbation, $a_s$ is the strong running coupling constant, and $h_n$ are the perturbative coefficients
\begin{equation} \label{UV}
h_n=h_{n,0}+h_{n,1}N_f+h_{n,2}N_f^2+\cdots + h_{n,n-1}N_f^{n-1}. 
\end{equation}
Here $N_f$ is the number of active flavor quarks. The $N_f$ terms come from the UV-divergent diagrams of the process, which depend dynamically on the virtuality of the fundamental quark and gluon subprocesses. We will show an optimal way to set the scales at each order of perturbation theory with the PMC method, in contrast to the conventional setting formalism, thus producing a relation that converges to the final value. Similarly, the inverse relation \cite{Steinhauser2016xkf,marquard2016ms,kataev2020multiloop,Kataev:2016jai,Kataev:2015gvt} is


\begin{equation} \label{zm1}
\overline{H}_{q}\left( \mu \right) =\frac{\overline{m}\left( \mu \right) }{M}=1+\sum_{n\geq 1}\overline{h}_{n}\left[a_{s}(\mu)\right] ^{n},
\end{equation}

with $\overline{h}_n$ the perturbative coefficients and, similar to Eq.\eqref{UV}, will be useful for determining the mass in the modified minimum subtraction scheme.

\section{Determination of the pole and $\overline{\text{MS}}$ mass for heavy quarks in the PMC setting}

In this section we present an improved analysis of the relation between the pole and $\overline{\text{MS}}$  quark masses using the PMC method. 
Before applying the PMC formalism, it will be necessary to obtain the dependence of the mass relation on the initial scale from the above mass relation with coefficients $N_f$ at a fixed scale. This can be done by using the scale dependence of the strong coupling constant up to the four-loop level of Eq.\eqref{relationalpha}, i.e.
\begin{eqnarray} \label{strong}
a_s(Q^*)&= & a_s(Q)-\beta_0 \ln\paren{\frac{Q^{*2}}{Q^2}}a_s^2(Q)\\
 \nonumber & & +\left[ \beta_0^2 \ln^2\paren{\frac{Q^{*2}}{Q^2}}-\beta_1 \ln\paren{\frac{Q^{*2}}{Q^2}}\right] a_s^3(Q)\\
\nonumber && + \left[ -\beta_0^3 \ln^3\paren{\frac{Q^{*2}}{Q^2}}+\frac{5}{2} \beta_0 \beta_1 \ln^2\paren{\frac{Q^{*2}}{Q^2}}\right. \\
\nonumber && \left. -\beta_2 \ln\paren{\frac{Q^{*2}}{Q^2}}\right] a_s^4(Q) + \mathcal{O}\paren{a_s^5}
\end{eqnarray}
where 
$Q^*$ and $Q$ are two arbitrary renormalization scales.

Choosing as the initial renormalization scale the pole mass of the heavy quark involved, the expression for the pole  mass in Eq.\eqref{cm1}, depending on the active quark flavor number $N_{f},$ can be written as
\begin{eqnarray} \label{Hrel}
H_{q} &=&1+h_{1,0}a_{s}\left( \mu _{r}^{init}\right) +\left(
h_{2,0}+h_{2,1}N_{f}\right) a_{s}^{2}\left( \mu _{r}^{init}\right) \\
& &+\left( h_{3,0}+h_{3,1}N_{f}+h_{3,2}N_{f}^{2}\right) a_{s}^{3}\left( \mu
_{r}^{init}\right) + \left( \right. h_{4,0}  \notag   \\
&&\left. +h_{4,1}N_{f}+h_{4,2}N_{f}^{2}+h_{4,3}N_{f}^{3}\right)
a_{s}^{4}\left( \mu _{r}^{init}\right) +\mathcal{O}\left( a_{s}^{5}\right) .
\notag
\end{eqnarray}%
The coefficients $h_{i,j}$, at four loop level, are given in Ref \cite{Steinhauser2016xkf}:

\begin{equation}
\begin{array}{lll}
h_{1,0}=1.333, &  & h_{2,0}=14.485, \\ 
h_{2,1}=-1,041, &  & h_{3,0}=217.903, \\ 
h_{3,1}=-27.961, &  & h_{3,2}=0,653, \\ 
h_{4,0}=4357.4\pm 1.64, &  & h_{4,1}=-\left( 834.548\pm 0.04\right), \\ 
h_{4,2}=-45.431, &  & h_{4,3}=-0.678.%
\end{array}
\end{equation}

The $\mathcal{R}_{\delta}$ scheme  \cite{mojaza2013systematic,brodsky2014systematic} reveals the special degeneracy of the coefficients in different perturbative orders of the series, ensuring the correspondence between the $N_{f}$ terms and the $\left\{\beta_{i}\right\}$ terms, through Eq.\eqref{strong}, order by order in the series. Considering what was mentioned earlier, from Eq.\eqref{Hrel}, the general form of the observable $H_q$ in terms of $\left\{\beta_{i}\right\}$ can be deduced given the structure of Eq.\eqref{Obs2}:
%
\begin{eqnarray} \label{Hbeta}
\nonumber H_{q} &=&1+r_{1,0}a_{s}\left( \mu _{r}^{init}\right) +\left( r_{2,0}+\beta
_{0}r_{2,1}\right) a_{s}^{2}\left( \mu _{r}^{init}\right)  \\
\nonumber &&+\left( r_{3,0}+\beta _{1}r_{2,1}+2\beta _{0}r_{3,1}+\beta
_{0}^{2}r_{3,2}\right) a_{s}^{3}\left( \mu _{r}^{init}\right)  \\
 &&+\left( r_{4,0}+\beta _{2}r_{2,1}+2\beta _{1}r_{3,1}+\frac{5}{2}\beta
_{0}\beta _{1}r_{3,2}\right.  \\
\nonumber && +3\beta _{0}r_{4,1}+3\beta _{0}^{2}r_{4,2}+\beta
_{0}^{3}r_{4,3}\bigg) a_{s}^{4}\left( \mu _{r}^{init}\right).
\end{eqnarray}

Here, for later convenience, we have transformed the series in $N_{f}$ into the required series $\left\lbrace \beta_i\right\rbrace $. The expressions for $\beta_0,\ \beta_1$ and $\beta_2$ of the above series were determined in \cite{baikov2017five}. The $r_{i,0}$ with $i=(1,\ldots,4)$ are scale-invariant conformal coefficients, and the $r_{i,j}$ with $1\leq j< i\leq 4$ are nonconformal coefficients that must be included in the coupling constant.

By applying PMC, we absorb into the coupling constant all the nonconformal coefficients that control the behavior of the coupling constant. The nonconformal terms are removed and the result of the pQCD series is transformed into a conformal mass ratio:

\begin{eqnarray}
H_{q}^{\text{PMC}} &=&1+r_{1,0}a_{s}\left( Q_{1}\right) +r_{2,0}a_{s}^{2}\left( Q_{2}\right)  \\
&&+r_{3,0}a_{s}^{3}\left( Q_{3}\right)
+r_{4,0}a_{s}^{3}\left( Q_{4}\right)  
\notag \\
&=&1+ 0.424307\alpha _{s}\left( Q_{1}\right)
- 0.272706\alpha _{s}^{2}\left( Q_{2}\right)   \notag \\
&&+ 3.27051\alpha _{s}^{3}\left( Q_{3}\right)
+ 1.84591\alpha _{s}^{4}\left( Q_{4}\right),   \notag
\end{eqnarray}

where the conformal coefficients are independent of the renormalization scale $\mu_r$. The resulting series becomes scheme independent up to the given order, and the ambiguity of the conventional renormalization scheme is removed. Here $Q_{k}\ \left(k=1,\ldots ,4\right) $ are the PMC scales, for the mass relation on shell-running mass PMC, calculated through Eq.\eqref{multiscale}.

The inverse relation, comparing the running mass with the pole mass, $\overline{H}_{q}$, conventionally determined at order $\mathcal{O}\left( a_{s}^{4 }\right)$ for a renormalization scale $\mu_{r}^{init}=M$, is

\begin{eqnarray}
\nonumber \overline{H}_{q} &=&1+\bar{h}_{1,0}a_{s}\left( \mu _{r}^{init}\right) +\left(
\bar{h}_{2,0}+\bar{h}_{2,1}N_{f}\right) a_{s}^{2}\left( \mu _{r}^{init}\right)  \\
 &&+\left( \bar{h}_{3,0}+\bar{h}_{3,1}N_{f}+\bar{h}_{3,2}N_{f}^{2}\right) a_{s}^{3}\left( \mu
_{r}^{init}\right)    \\
\nonumber &&+\left( \bar{h}_{4,0}+\bar{h}_{4,1}N_{f}+\bar{h}_{4,2}N_{f}^{2}+\bar{h}_{4,3}N_{f}^{3}\right)
a_{s}^{4}\left( \mu _{r}^{init}\right) ,
\label{Hgeneral}
\end{eqnarray}
where the coefficients are set out in \cite{Steinhauser2016xkf}:
\begin{equation}
\begin{array}{lll}
\bar{h}_{1,0}=-1.333, &  & \bar{h}_{2,0}=-15.374, \\ 
\bar{h}_{2,1}=1.041, &  & \bar{h}_{3,0}=-226.283, \\ 
\bar{h}_{3,1}=28.229, &  & \bar{h}_{3,2}=-0.653, \\ 
\bar{h}_{4,0}=-4455.25\pm 1.64, &  & \bar{h}_{4,1}=845.941\pm 0.04, \\ 
\bar{h}_{4,2}=-45.517, &  & \bar{h}_{4,3}=0.678.%
\end{array}%
\end{equation}

Performing the PMC adjustment one arrives at the following conformal series:
\begin{eqnarray}
\overline{H}_{q}^{\text{PMC}} &=&1+\overline{r}_{1,0}a_{s}\left( \overline{Q}_{1}\right) +\overline{r}_{2,0}a_{s}^{2}\left( \overline{Q}_{2}\right)  \\
&&+\overline{r}_{3,0}a_{s}^{3}\left( \overline{Q}_{3}\right)
+\overline{r}_{4,0}a_{s}^{3}\left( \overline{Q}_{4}\right),  
\notag \\
&=&1- 0.424307\alpha _{s}\left( \overline{Q}_{1}\right)
 0.182631 \alpha _{s}^{2}\left( \overline{Q}_{2}\right)   \notag \\
&&- 3.39816\alpha _{s}^{3}\left( \overline{Q}_{3}\right)
- 1.60253\alpha _{s}^{4}\left( \overline{Q}_{4}\right)  , \notag
\end{eqnarray}
where $\overline{Q}_k\ \left( k=1,\ldots ,4\right)$ are the PMC scales at each order of perturbation, determined by eqs. \eqref{multiscale}, \eqref{rrr}, \eqref{rrr2} and (\ref{rrr3}), using the conformal coefficients $\overline{r}_{i,0}$ and the nonconformal coefficients $\overline{r}_{i,j}$ of the inverse relation.\\

The results presented in this section are $H_q^{\text{PMC}}$ and $\overline{H}_q^{\text{PMC}}$, which initially come from the series determined in the on-shell and $\overline{\text{MS}}$ schemes.  After following the PMC formalism, they become scheme independent. The initial schemes are used as a way to determine the generalized relation for the pole and running masses, and using this we can determine the mass relations of the heavy quarks $M_q^{\text{PMC}}$ and their corresponding mass differences $\Delta M_q^{\text{PMC}}$ in the PMC setting.

\section{Numerical results and discussions}
\label{sec:num}

For the numerical calculations we will use the central value for the current masses of the $c$, $b$ and $t$ quarks from PDG-$2022$ \cite{particle2022review}, namely 

\begin{equation}\label{inputmsbar1}
\begin{aligned}
\overline{m}_c\paren{\overline{m}_c^2} = &\ 1. 27 \pm 0.02 \ \text{GeV}, \  \alpha _{s}\left( \overline{m}_{c}\right) =0.38\pm 0.03,\\
\overline{m}_b\paren{\overline{m}_b^2}= &\ 4.18_{-0.03}^{+0. 04}\ \text{GeV}, \ \alpha _{s}\left( \overline{m}_{b}\right) =0.233\pm 0.008,
\end{aligned}
\end{equation}
and from cross section measurements 
\begin{equation}\label{inputmsbar2}
\overline{m}_t\paren{\overline{m}_t^2}=162.5_{-1.5}^{+2. 1} \ \text{GeV}, \ \alpha _{s}\left( \overline{m}_{t}\right) =0.1083.
\end{equation}
As input parameters we use the following central values for the pole masses of the quarks from PDG-$2022$ \cite{particle2022review} to determine the $\overline{\text{MS}}$ masses
\begin{equation}\label{inputpole}
\begin{aligned}
M_c = & \ 1.67 \pm 0.07\ \text{GeV},\  M_b= 4.78\pm 0.06 \ \text{GeV}\\
M_t= & 172.5 \pm 0.7\ \text{GeV},\ \alpha_{s}\left( M_{Z}\right) =0.1181.
\end{aligned}
\end{equation}

\subsection{Charm quark mass relations}
The determination of the charm quark mass plays an essential role in the estimation of the CKM parameters.  The relation of its on-shell and running masses with QCD corrections $\paren{\alpha_s}$ have been obtained up to  three loops \cite{ hoang2006charm} and four loops \cite{kataev2020multiloop}. The results $H_c$ are presented in Table~\ref{tableC}, where $H_{c}^{\left( i\right) }\ $represents the approximate relation for each perturbative order with $i=$LO, NLO, etc., where we have fixed $\mu_r=\mu^{init}=2$ GeV for the conventional scale setting, and in the PMC scaling we take $\mu^{init}=2$ GeV for the initial scale.

For the relation $H_{c}$ we need to introduce four PMC scales. As indicated in the previous section, Eq.\eqref{multiscale}, since there are not enough $\beta$ terms to determine its optimal value from the last scale, we set $Q_{4}^{\left( c\right) }$ equal to the last determined scale of the process. 

The PMC multi-scales are
\begin{table*}[htbp]
\begin{tabular}{ccccccccccc}
\hline
 &$\quad$ & $H_{c}^{\left( 1\right) }$ & $H_{c}^{\left( 2\right) }$ & $H_{c}^{\left(
3\right) }$ & $H_{c}^{\left( 4\right) }$ & $\quad$ &$\overline{H}_{c}^{\left( 1\right) }$ & $\overline{H}_{c}^{\left( 2\right) }$ & $\overline{H}_{c}^{\left(
3\right) }$ & $\overline{H}_{c}^{\left( 4\right) }$  \\ \hline
Conv. & & $1.16124$ &  $1.31224$& $ 1.51842 $& $1.88291$
& & $0.949974$ & $0.934186$ & $0.927642$ & $0.924158$ \\
PMCs & & $1.10309$ & $1.08699$&  $1.13389$&  $1.14032$ &  &$0.947721$ & $0.950494$ & $0.944138$ & $0.943769$\\
 PMC & & $1.24077$ & $1.12431$ & $1.92046$ & $2.20104$ &  &$0.944683$ & $0.949485$ & $0.941559$ & $ 0.941064$ \\ \hline
\end{tabular}
\caption{Numerical results for the c-quark $H_c$ mass ratio and inverse $\overline{H}_c$ ratio, under conventional (Conv.) scale setting, PMCs and PMC, respectively, corrected up to four-loop. $\mu_r^{init}=2$ GeV (and $\mu_r^{init}=M_z$ \cite{particle2022review}).}
\label{tableC}
\end{table*}
\begin{equation}
\begin{array}{c}
Q_{1}^{\left( c\right) }=1.136\ \text{GeV, }Q_{2}^{\left( c\right)
}=0.950\ \text{GeV}, \\ 
Q_{3}^{\left( c\right) }=1.006\ \text{GeV},
\end{array}
\end{equation}

which are different and smaller than $\mu^{init}$, showing that they are governed by a different series $\left\lbrace \beta_{i}\right\rbrace $ at each perturbative order.
The QCD perturbative corrections to the pole-running charm-quark mass ratio are
\begin{equation}
H_c^{\text{PMC}}=1+0.2408-0.1165+0.7962+0.2806.
\end{equation}

An alternative approach is to use the PMC single scale setting (PMCs), as this helps reduce the dependence on a residual scale when it becomes significant. The idea is to set a single effective scale.  The details of the procedure are illustrated in \cite{singlescale1}. In our case, the single scale is determined from the following perturbative relation, fixed to order next-to-next-to-leading log (N$^2$LL) as
\begin{equation}
\ln \left( \frac{Q_{s}^{\left( c\right) }}{Q}\right)
^{2}=-1.171+12.56 a_s(Q)+104.8 a_s^2(Q),
\end{equation}
where $Q=\mu_r^{init}$ and the coefficients are determined according to the formulas in the Appendix. \ref{single}.
\begin{figure}[H]
\centering
\includegraphics[scale=.6]{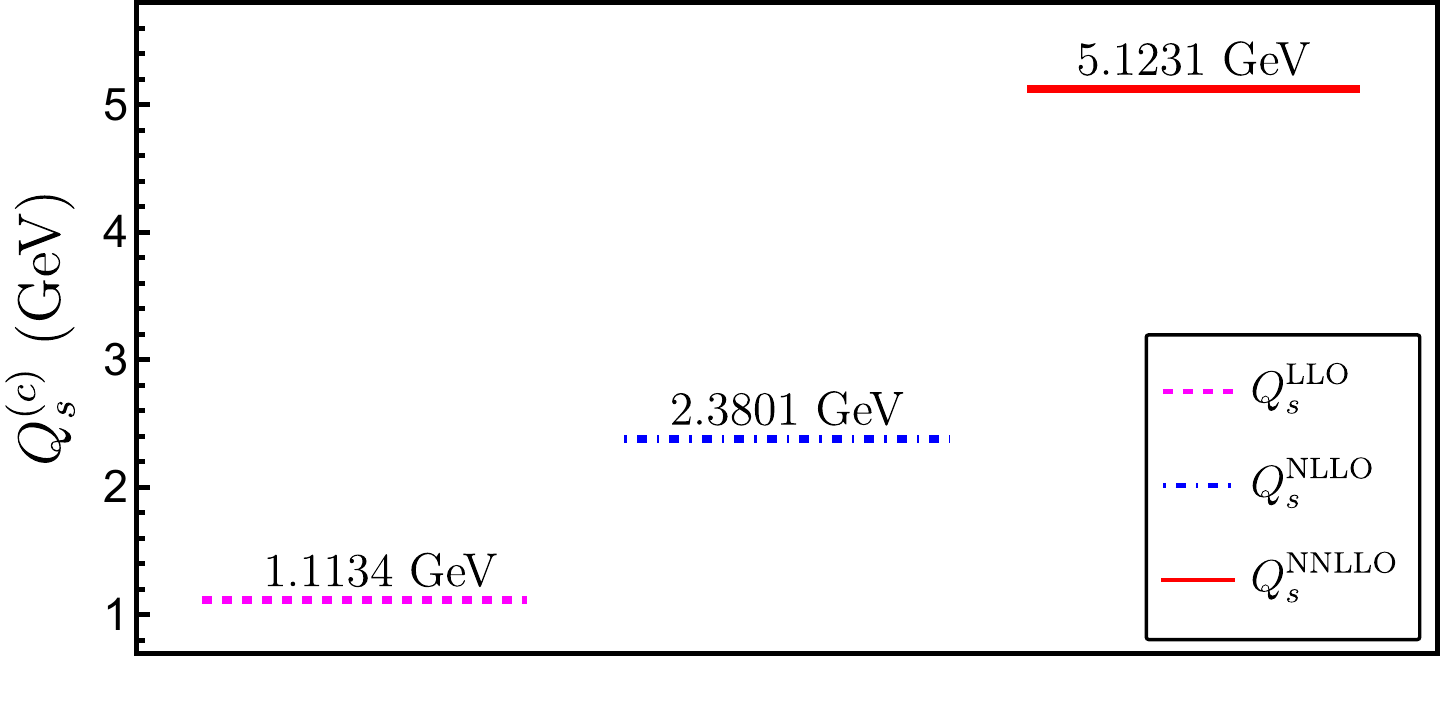}
\caption{Single-scale PMC values for $H_c$, determined up to NNLL. $\mu_r^{init}=2$ GeV.}
\label{figcharm}
\end{figure}

The PMC single scale values are shown in Fig.~\ref{figcharm}, up to order N$^2$LL $Q_s^{(c)}=1.1134$ GeV. The scales at different orders are $Q_{s}^{\left( c\right) ,\text{LL}}<Q_{s}^{\left( c\right),\text{NLL}} < Q_{s}^{\left( c\right),\text{N}^{2}\text{LL}}$. 

The best result for the PMC pole mass of the charm quarks is obtained by performing the renormalization scale fit using PMCs, since this gives a faster convergence of the observables. The values for each approximation are
\begin{equation}
\begin{array}{l}
\left. M_{c}^{\text{PMC}}\right\vert _{\text{NLO}}=1.40_{-0.02}^{+0.05}\ \text{GeV}, \\ 
\left. M_{c}^{\text{PMC}}\right\vert _{\text{N}^{2}\text{LO}}=1.38_{-0.02}^{+0.03}\ \text{GeV}, \\ 
\left. M_{c}^{\text{PMC}}\right\vert _{\text{N}^{3}\text{LO}}=1.44_{-0.05}^{+0.14}\ \text{GeV}, \\ 
\left. M_{c}^{\text{PMC}}\right\vert _{\text{N}^{4}\text{LO}}=1.45_{-0.05}^{+0.17}\ \text{GeV}.
\end{array}
\label{eq:mcpolePMCs}
\end{equation}
The perturbative PMC Pole mass $M_c^{\text{PMC}}$ contributes to the central value in \cite{particle2022review}, and is comparable to the results shown in refs. \cite{kataev2020multiloop,marquard2016ms}.

The uncertainties in Eq.~\eqref{eq:mcpolePMCs} are determined by considering the variation of the single scale in the orders $\text{N}^2\text{LL}$ and $\text{NLL}$, i.e. $ Q_s^{(c),\text{N}^2\text{LL}}\pm \lvert Q_s^{(c),\text{N}^2\text{LL}} -Q_s^{(c),\text{NLL}} \rvert $.
 
To quantify the theoretical prediction, we will estimate the error by defining as
\begin{equation}
E\paren{M_q}= \frac{\lvert M_q^{\text{data}}-M_q \rvert}{1\ \text{GeV}},
\end{equation}
which compares the obtained numerical result with the value extracted from the experimental or phenomenological data, i.e. for the charm quark the conventional setting $E(M_q^{\text{Conv.}})\simeq 0.7$ and our result $E(M_q^{\text{PMCs}})\simeq 0.2$.

The values of the inverse relation, Eq.\eqref{zm1}, of the charm quark $\overline{H}_c$ are shown in Table \ref{tableC} from PMC and PMCs, compared with the conventional scale setting. When the PMC procedure is applied, the values of the masses shown in Table \ref{c-mass} are obtained. Finally, the difference between the PMC pole mass and the running $\Delta M_c=M_c-\overline{m}_c$ charm quark approaches a better result with a difference of $\Delta M_c^{\text{PMC}} \sim 100$ MeV.

\begin{table}
\centering
\begin{tabular}{ccccccc}
\hline
 & & $M_c$ (GeV) & & $\overline{m}_c$ (GeV)& & $\left\vert \Delta M_c \right \vert $ (GeV)\\ \hline
Conv.& &$2.39\pm 0.04$ & & $1.54^{+0.02}_{-0.03}$ & & $0.847947$ \\
PMCs & &$1.45^{+0.17}_{-0.05}$ & &$1.576^{+0.0012}_{-0.0013}$ & & $0.108979$ \\
PMC & &$2.79 \pm 0.04$ & & $1.57\pm 0.07$ & &$0.939072$\\ \hline
\end{tabular}
\caption{
Values for the pole mass whit input Eq.\eqref{inputmsbar1}, running mass whit input Eq.\eqref{inputpole} and the difference between them for the charm quark, using the conventional setting (Conv.), PMCs, and PMC.}
\label{c-mass}
\end{table}

\subsection{Bottom quark mass relations}

The $b$ quark plays an important role in modern particle physics, and a precise knowledge of its mass parameter is necessary for accurate theoretical predictions, e.g. a precise bottom quark mass is required for $b$ meson decay calculations, which is often proportional to the fifth power of the quark mass \cite{marquard2015quark}.
\begin{table*}[htbp]
\centering
\begin{tabular}{ccccccccccc}
\hline
 & $\quad$& $H_{b}^{\left( 1\right) }$ & $H_{b}^{\left( 2\right) }$ & $H_{b}^{\left(
3\right) }$ & $H_{b}^{\left( 4\right) }$ &$\quad$& $\overline{H}_{b}^{\left( 1\right) }$ & $\overline{H}_{b}^{\left( 2\right) }$ & $\overline{H}_{b}^{\left(
3\right) }$ & $\overline{H}_{b}^{\left( 4\right) }$  \\ \hline
Conv. & & $1.09462$ & $1.14138$ & $1.17515$ & $1.20652$ & & $0.949889$ & $0.935519$ & $0.930128$ & $0.927575$  \\
PMCs& & $1.09197$ & $1.07916$ & $1.11247$ & $1.11654$ & &$0.947965$ & $0.950712$ & $0.944444$ & $0.944082$ \\
 PMC & & $1.11393$ & $1.08096$ & $1.15252$ & $1.16382$ & & $0.94520$ & $ 0.949855$ & $0.942191$ & $0.941717$ \\ \hline
\end{tabular}
\caption{Numerical results for the b-quark $H_b$ mass ratio and the inverse $\overline{H}_b$ ratio, under conventional (Conv.) scale setting, PMCs and PMCs corrected up to four loops, respectively. $\mu_r^{init}=2$ GeV (and $\mu_r^{init}=M_z$).}
\label{tableB}
\end{table*}

\begin{figure}[]
\centering
\includegraphics[scale=0.6]{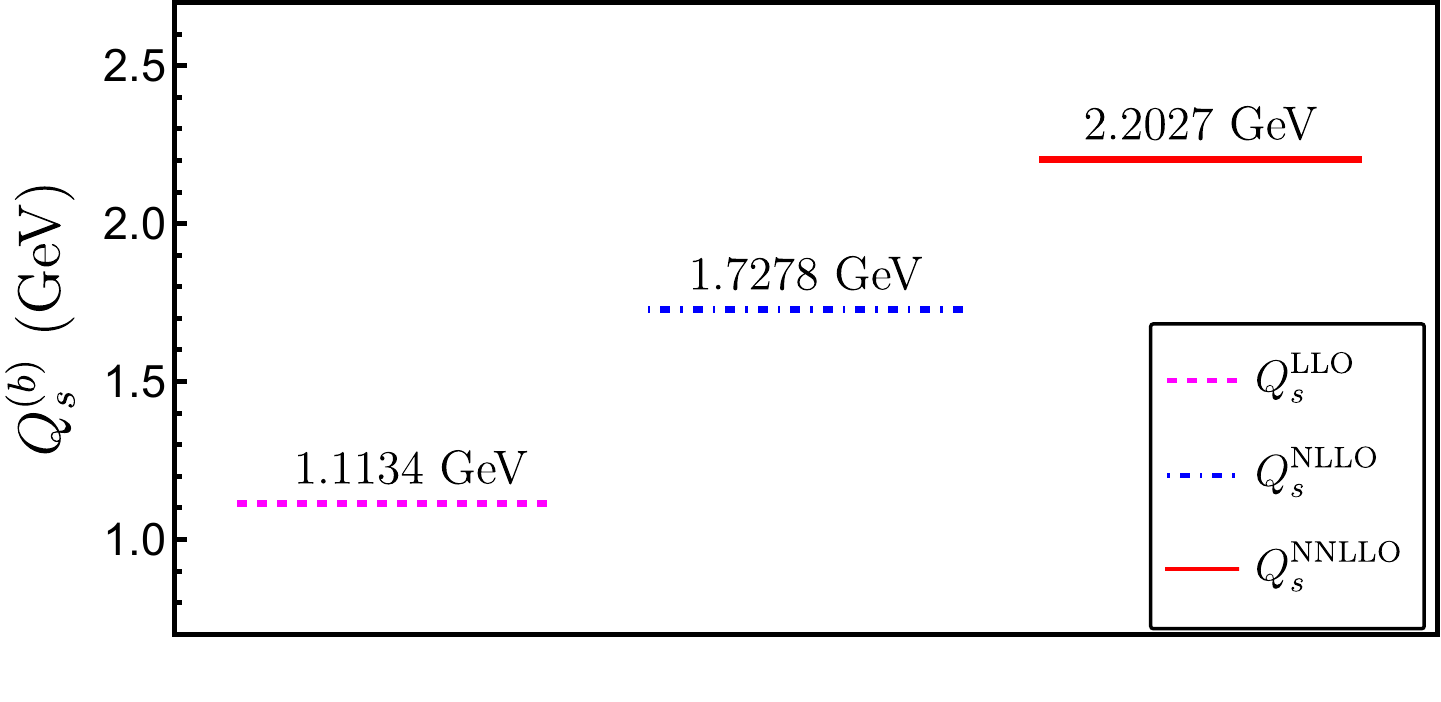}
\caption{Single-scale PMC value for $H_b$, determined up to NNLL. $\mu_r^{init}=2$ GeV.}
\label{figbottom}
\end{figure}

Perturbative bottom-quark mass relations with QCD corrections are described up to  three loops in ref. \cite{chetyrkin2000relation} and four loops, in refs. \cite{melnikov2000three, Steinhauser2016xkf, kataev2020multiloop}.  Using the PMC formalism for the $b$ quark, the following numerical series are obtained
\begin{equation}
H_b^{\text{PMC}}=1+0.1140-0.03297+0.07156+0.01129,
\end{equation}
in contrast to the displayed value $\mathcal{O}\paren{\alpha_s^3}$ in \cite{particle2022review} and $\mathcal{O}\paren{\alpha_s^4}$ in \cite{marquard2015quark,kataev2020multiloop}, where the PMC multi-scale adjustment provides the following scales

\begin{equation}\label{escalasb}
\begin{array}{c}
Q_{1}^{\left( b\right) }=1.129\ \text{GeV, }Q_{2}^{\left( b\right)
}=0.5682\ \text{GeV}, \\ 
Q_{3}^{\left( b\right) }=1.006\ \text{GeV,}
\end{array}
\end{equation}
where $Q_4^{(b)}$ is set equal to the last determined scale in the main order. Optionally, in the single scale fit (PMCs), for $Q_s^{(b),\text{NNLLO}}$ the mass ratio $H_b$ is presented in Table \ref{tableB}, where an initial scale choice of $2$ GeV is taken for the conventional scale setting, compared to the multi-scale in Eq.\eqref{escalasb} and the single scale setting, with values presented in Fig.\ref{figbottom}, given by
\begin{equation}
\ln \left( \frac{Q_{s}^{(b) }}{Q}\right)
^{2}=-1.1714+12.38 a_s(Q)+ 96.38 a_s^2(Q),
\end{equation}
to order N$^2$LL.

Our results for the PMC pole mass of the bottom quark, presented in different orders, are
\begin{equation}
\begin{array}{l}
\left. M_{b}^{\text{PMC}}\right\vert _{\text{NLO}}=4.66_{-0.02}^{+0.03}\ \text{GeV}, \\ 
\left. M_{b}^{\text{PMC}}\right\vert _{\text{N}^{2}\text{LO}}=4.52_{-0.02}^{+0.03}\ \text{GeV} ,\\ 
\left. M_{b}^{\text{PMC}}\right\vert _{\text{N}^{3}\text{LO}}=4.82_{-0.02}^{+0.04}\ \text{GeV}, \\ 
\left. M_{b}^{\text{PMC}}\right\vert _{\text{N}^{4}\text{LO}}=4.86_{-0.02}^{+0.03}\ \text{GeV}.
\end{array}
\label{eq:mbpolePMCs}
\end{equation}


where the uncertainties are determined by considering the variation of the multi-scale $Q_k^{(b)}$ (with $k=1,2,3$) in the conventional range in a truncation procedure. It is important to point out that the mass of the bottom, not being an observable, has an uncertainty associated with $\Lambda_{\text{QCD}}$ of the order of $0.3$ GeV, which is always present and cannot be suppressed by the OPE procedure, in this case this uncertainty is larger than the one delivered by the PMC prescription, so it represents the main uncertainty, then the central values delivered by the PMC are affected by this value. These results show that the QCD correction is $14.5\%$ of the total value. The value obtained for the PMC pole mass, $M_b^{\text{PMC}}=4.84$ GeV, considering the QCD four-loop corrections, with $E\paren{M_b^\text{PMC}} \simeq 0.08$, is much closer to the central value of PDG($2022$), $M_{b}=4.78\pm 0.06$ GeV \cite{particle2022review}, than the value obtained with the conventional scale setting $M_b^{\text{Conv.}}$, $E\paren{M_b^\text{Conv.}} \simeq 0.26$, which contains the ambiguity of the renormalization scheme. On the other hand, the value obtained by the second PMC approach is very close to the central value. The numerical correction to order $\mathcal{O}\paren{\alpha_s^4}$ for the inverse relation for the quark bottom $\overline{H}_b^{\text{PMC}}$ is presented in Table \ref{tableB}.

\begin{figure}[]
\centering
\includegraphics[scale=.57]{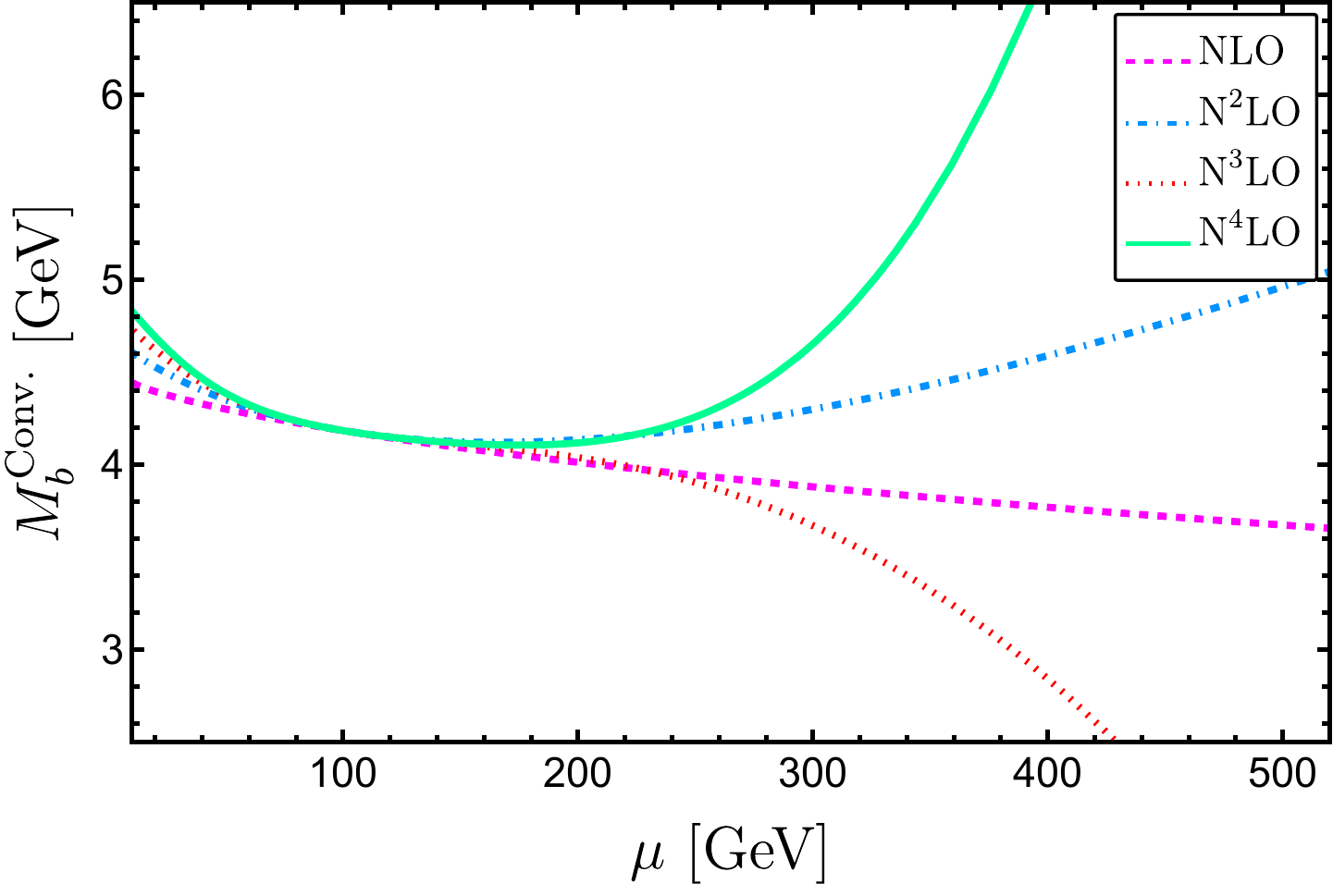}
\caption{Plot one-loop (dashed line), two-loop (dotdashed line), three-loop (dotted line) and four-loop (solid line) QCD correction to $M_b$ to convenctional scale setting. For $N_f=5$ and $\mu_r^{init}=2\ \text{GeV}$.}
\label{plotbottomconv}
\end{figure}

\begin{figure}[]
\centering
\subfigure[]{\includegraphics[scale=.57]{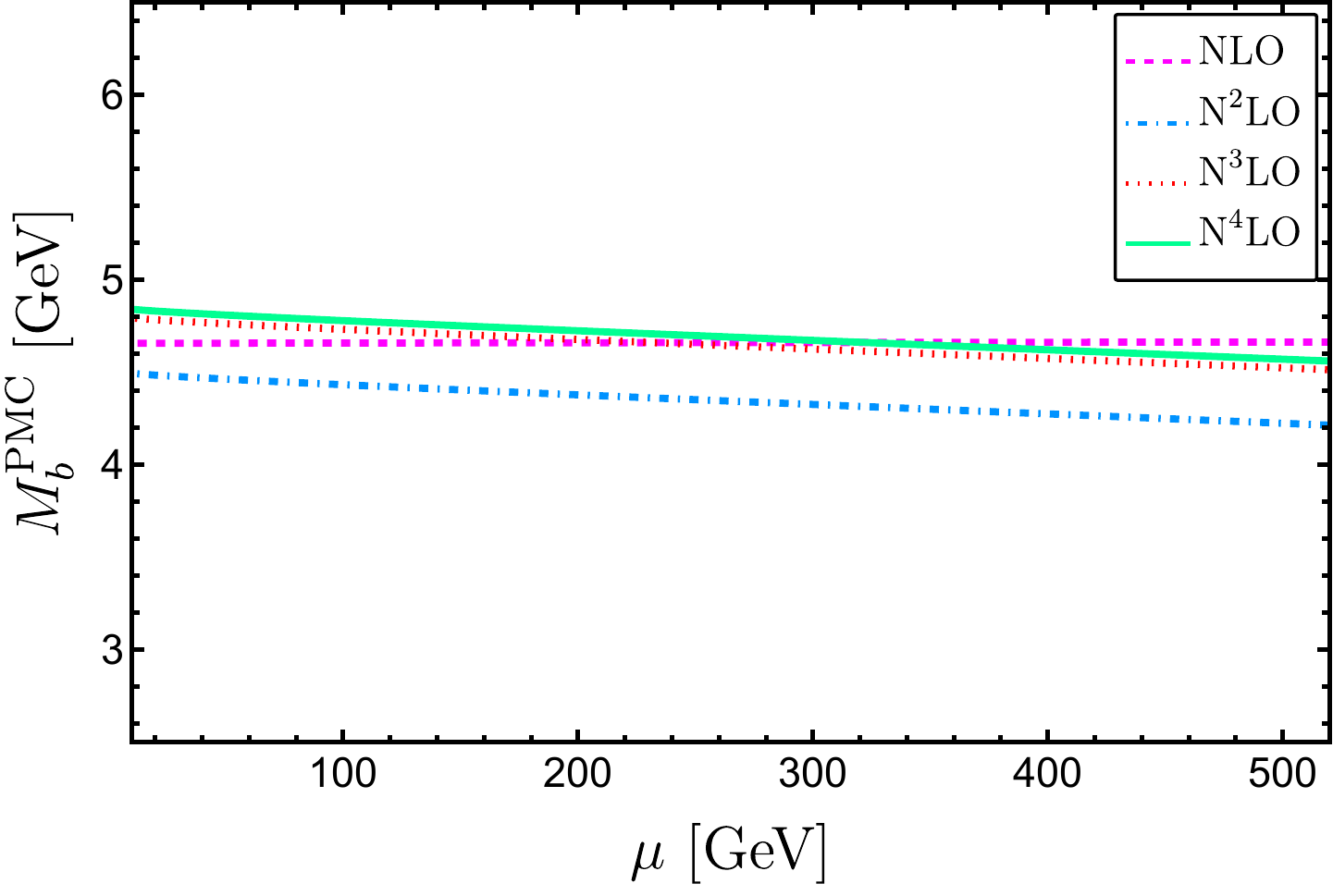}}
\subfigure[]{\includegraphics[scale=.57]{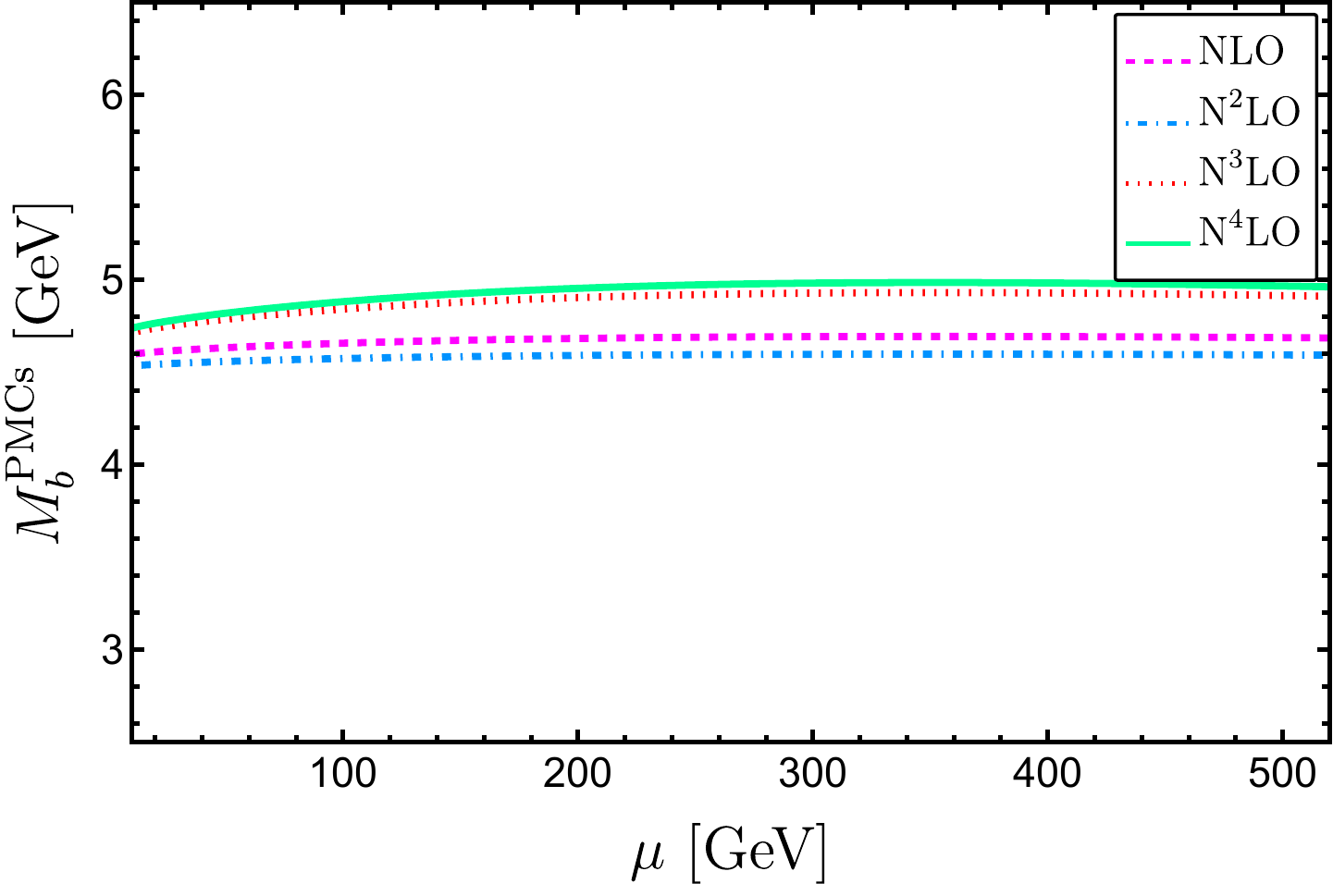}}
\caption{Plot one-loop (dashed line), two-loop (dotdashed line), three-loop (dotted line) and four-loop (solid line) QCD correction to $M_b$ in the setting (a) PMC multi-scale and (b) PMCs single-scale. For $\mu_r^{init}=2\ \text{GeV}$.}
\label{plotbottomPMC}
\end{figure}

In the conventional scale setting, the dependence on the renormalization scale significantly alters the value of the $b$ quark pole mass, $M_b$, shown in Fig.~\ref{plotbottomconv}  as a function of the renormalization scale. The large dependence is mainly due to the arbitrary choice of the renormalization scale. Due to the perturbative nature of the theory, the mass determination undergoes a poor convergence by choosing the scale as the transferred momentum, introducing relevant uncertainties in the predictions. 
On the other hand, in the scale setting, the principle of maximum conformality contributes a systematic way to determine the renormalization scale, improving the accuracy of the bottom quark mass determination. In this approach, it is possible to absorb the large logarithms of the QCD corrections and obtain a fully conformal series. In Fig.~\ref{plotbottomPMC} the bottom quark mass in the PMC is plotted as a function of the renormalization scale. A comparison of the graphical results in Fig.~\ref{plotbottomconv}-\ref{plotbottomPMC} shows a significant improvement in convergence and a reduction in uncertainty at high energies. In Fig.~\ref{plotbottomPMC} \textcolor{red}{(a)} the calculations are performed with the PMC multi-scale setting. In contrast, in Fig~\ref{plotbottomPMC} \textcolor{red}{(b)} uses the single-scale setting of the PMCs, which is a good alternative to simplify the calculations by reducing the number of parameters.

In Table \ref{b-mass} the numerical results for the PMC mass of the $b$ quark in both relations are shown. The difference in the PMC mass for the $b$ quarks is reduced by $0.34$ GeV.

\begin{table}[]
\centering
\begin{tabular}{ccccccc}
\hline
 & & $M_b$ (GeV) & & $\overline{m}_b$ (GeV)& & $\left\vert \Delta M_b \right \vert $ (GeV)\\ \hline
Conv.& &$5.04^{+0.86}_{-0.51}$ & & $4.43^{+0.05}_{-0.06}$ & & $0.608598$ \\
PMCs & &$4.67^{+0.06}_{-0.04}$ & & $4.512^{+0.003}_{-0.006}$ & & $0.155076$ \\
PMC & &$4.86^{+0.03}_{-0.02}$ & & $4.50\pm 0.06$ & &$0.363357$\\ \hline
\end{tabular}
\caption{Obtained values for the pole mass whit input Eq.\eqref{inputmsbar1}, running mass whit input Eq.\eqref{inputpole} and the difference between them for the b-quark, using the conventional setting (Conv.), PMCs, and PMC.}
\label{b-mass}
\end{table}

\subsection{Top quark mass relations}
\label{subsection:top}

\begin{figure}[H]
\centering
\includegraphics[scale=0.6]{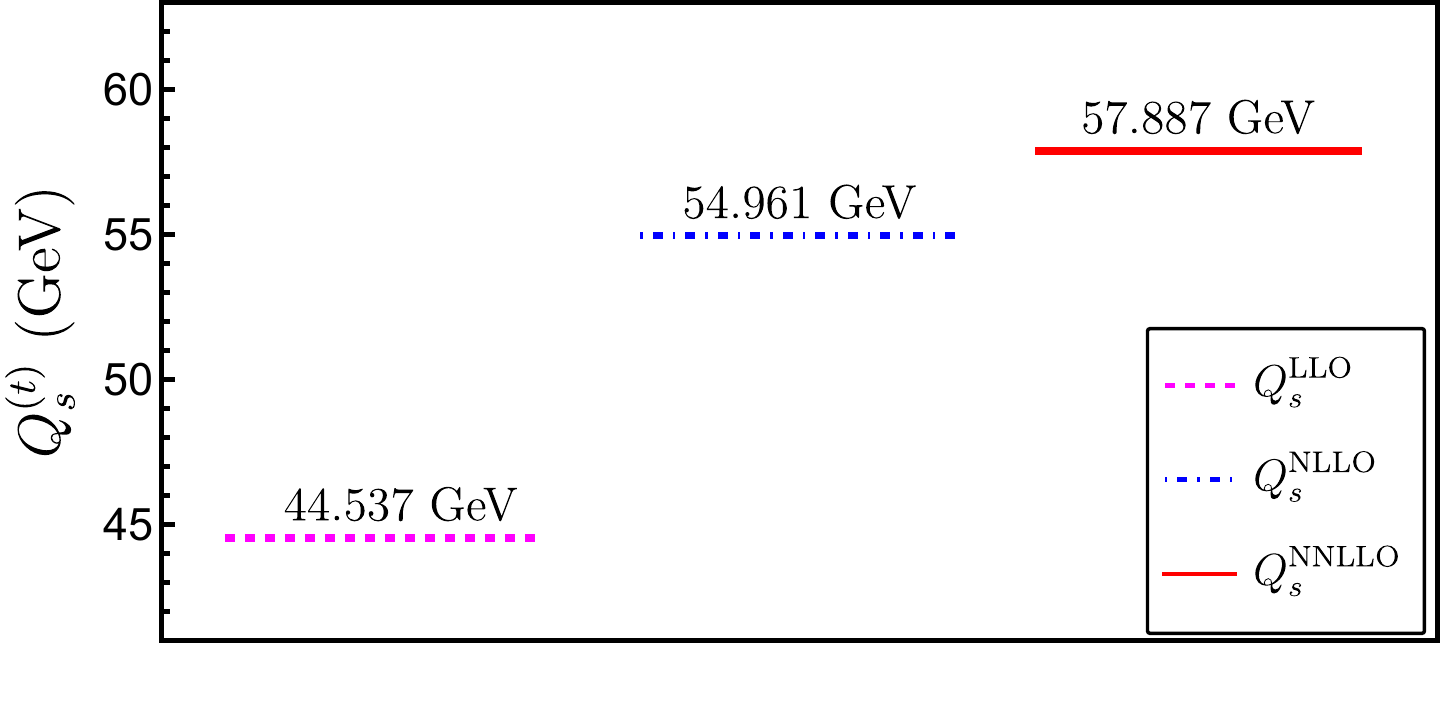}
\caption{Single-scale PMC value for $H_t$, determined up to NNLL. $\mu_r^{init}=80$ GeV.}
\label{figtop}
\end{figure}

The top quark mass is an essential theoretical parameter extracted from Tevatron and LHC experimental data in \cite{particle2022review,hoang2020top}. The renormalized mass of the top quark, in the pole-running mass relation with QCD corrections up to $\mathcal{O}\paren{\alpha_s^3}$ and $\mathcal{O}\paren{\alpha_s^4}$, has been determined in \cite{chetyrkin2000relation} and \cite{marquard2015quark, marquard2016ms, Steinhauser2016xkf, kataev2020multiloop} and with mixed QCD-EW corrections $\paren {\alpha \alpha_s}$ in \cite{jegerlehner2003alpha, kataev2022notes}. The pole mass is sensitive to small momenta, implying a rather large perturbative correction; thus, truncating the series in a finite order inevitably makes it scale scheme dependent. The Principle of Maximum Conformality greatly improves the mass ratios for the top quark, giving results independent of the renormalization scheme. The PMC mass ratio (and the inverse ratio) at different orders of Perturbative Theory (PT) are shown in Tab.\ref{tableT}, for conventional scaling, PMC and PMCs, taking an initial scale of $80$ GeV. The value of the ratio under the PMC formalism, up to $\mathcal{O} \paren{\alpha_s^4}$, is
\begin{equation}
H_t^{\text{PMC}}=1+\paren{49.5-5.00+5.43+0.363}\times 10^{-3},
\end{equation}
This is a more convergent result than Eq.\eqref{Hbeta} for the conventional case. The effective scales found are
\begin{equation}\label{escalastop}
\begin{array}{c}
Q_{1}^{\left( t\right) }=44.86\ \text{GeV, }Q_{2}^{\left( t\right)
}=16.24\ \text{GeV}, \\ 
Q_{3}^{\left( t\right) }=40.25\ \text{GeV.}
\end{array}
\end{equation}	
These values are independent of the initial renormalization scale, except for some residual dependence on the truncated $\beta$ function, which is less than the quoted precision. 
Table.\ref{tableT} shows the results for PMCs, where the unique scale is shown in Figure \ref{figtop}, up to order NNLL (next-to-next-to-leading logarithm). The results for the PMC pole mass of the top quark, for different perturbation orders, are shown below:
\begin{equation}
\begin{array}{l}
\left. M_{t}^{\text{PMC}}\right\vert _{\text{NLO}}=172.1\pm 0.6\ \text{GeV} ,\\ 
\left. M_{t}^{\text{PMC}}\right\vert _{\text{N}^{2}\text{LO}}=171.3 \pm 0.6\ \text{GeV}, \\ 
\left. M_{t}^{\text{PMC}}\right\vert _{\text{N}^{3}\text{LO}}=172.2\pm 0.6\ \text{GeV}, \\ 
\left. M_{t}^{\text{PMC}}\right\vert _{\text{N}^{4}\text{LO}}=172.3 \pm 0.6\ \text{GeV}.
\end{array}
\end{equation}
The PMC pole Mass of the Top Quark 
is a significant improvement over the conventional determination. The current central value is $172.5$ GeV \cite{particle2022review} and contributes to the value determined in the process of top quark pair production at the hadron colliders in \cite{wang2018precise}.

\begin{figure}[]
\centering
\subfigure[]{\includegraphics[scale=.57]{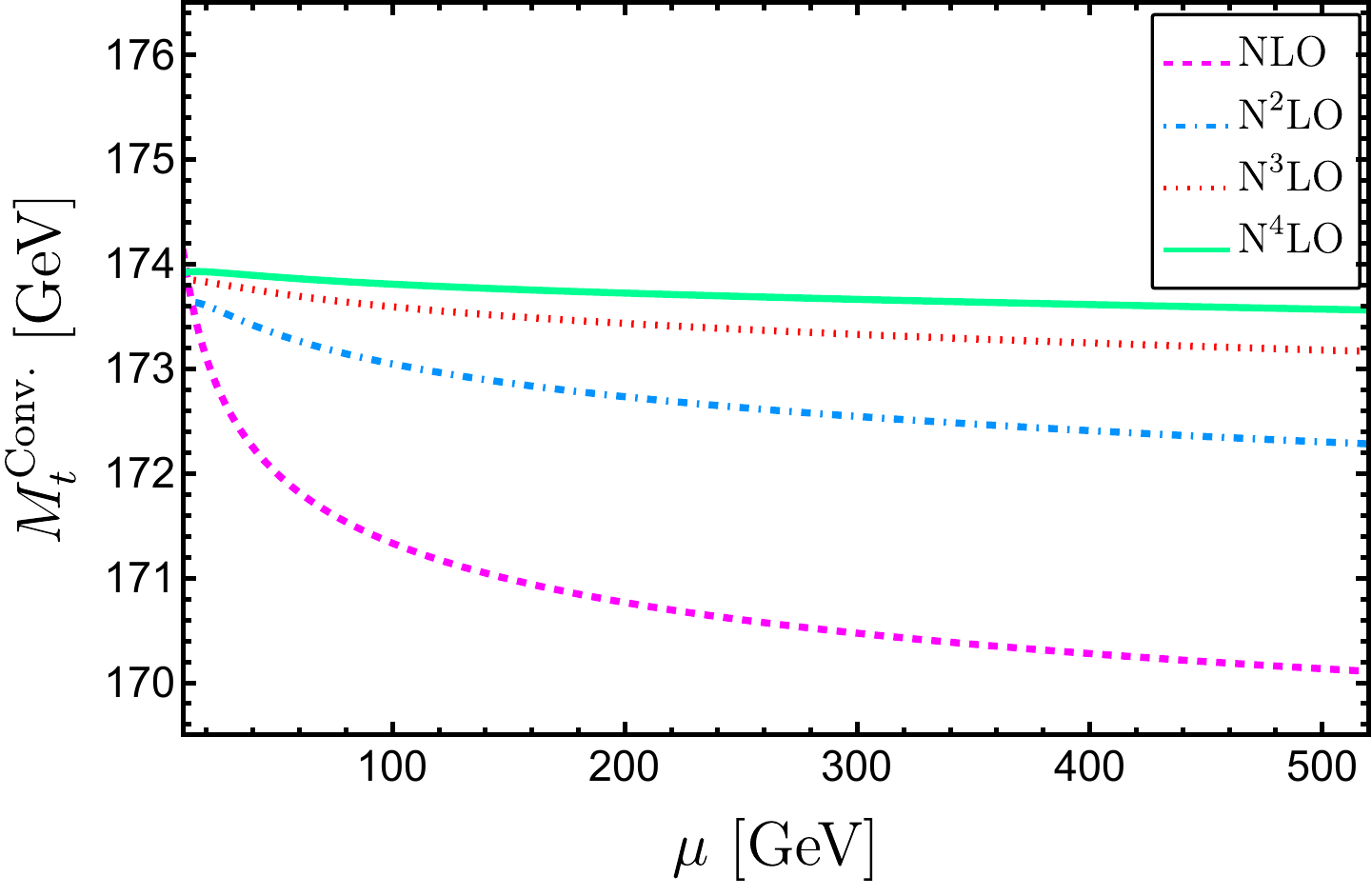}}
\caption{Plot one-loop (dashed line), two-loop (dotted line), three-loop (dotted line), and four-loop (solid line) QCD corrections to $M_t$ at the conventional scale setting. For $N_f=6$ and $\mu_r^{init}=80\ \text{GeV}$.}
\label{plottopconv}
\end{figure}

\begin{figure}[]
\centering
\subfigure[]{\includegraphics[scale=.57]{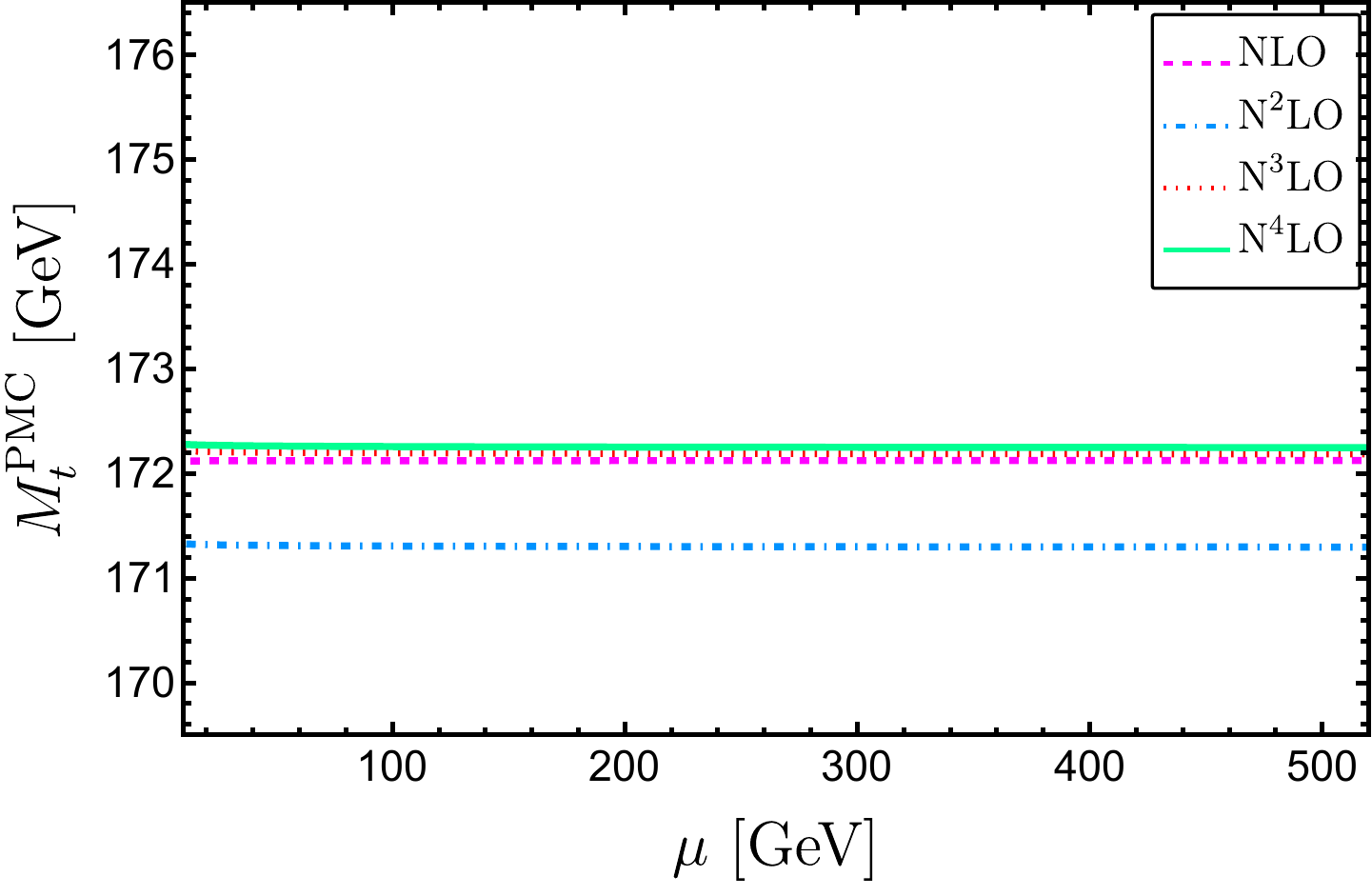}}
\subfigure[]{\includegraphics[scale=.57]{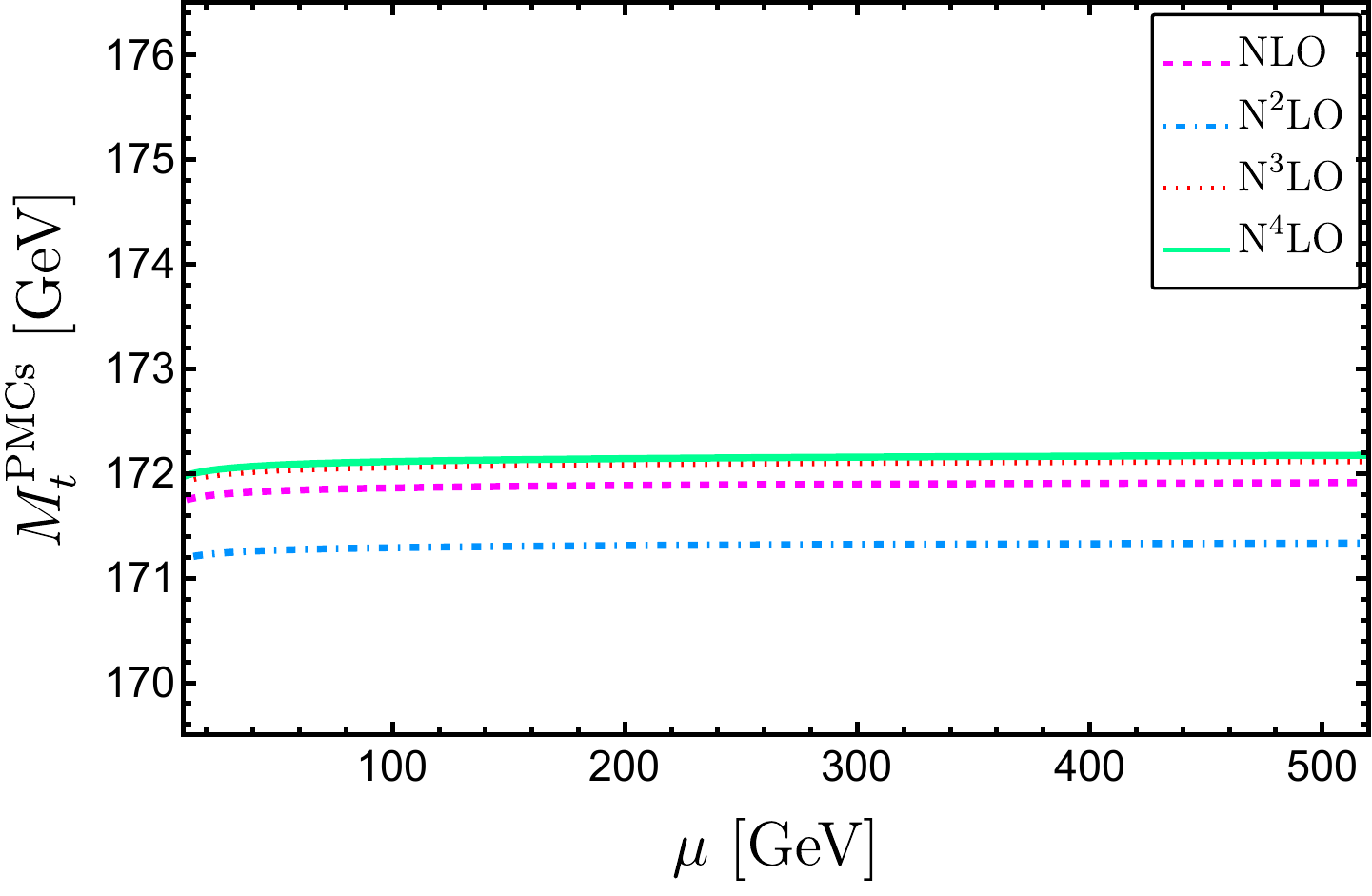}}\caption{Plots of one-loop (dashed line), two-loop (dotted dashed line), three-loop (dotted line) and four-loop (solid line) QCD corrections to the top-mass quark $M_t$ (a) PMC multi-scale setting, and (b) PMC single-scale setting. For $\mu_r^{init}=80\ \text{GeV}$.}
\label{plottopPMC}
\end{figure}

In Fig.~\ref{plottopconv} we show the dependence of the renormalization scale in the conventional fit for the top quark mass value $M_t$, up to four loops, using the expression Eq.~\eqref{Hbeta}, with an initial renormalization scale of $80$ GeV. The numerical case for the determination of the top quark mass after the multi-scale PMC and single-scale PMCs fits is shown in Fig.~\ref{plottopPMC}, where it can be observed that the dependence on the renormalization scale is suppressed. This is due to the systematic determination of the PMC effective scales, which improves the uncertainty compared to the conventional case. In Fig.~\ref{plottopPMC}(a) we show the multi-scale case, and in Fig.~\ref{plottopPMC}(b) we show the single-scale case.

The obtained result for the top quark PMC/running mass is shown in Table~\ref{t-mass}, which also shows an improvement. The top quark PMC mass correction can be used as a good fit parameter in the slant corrections of $\Delta \rho$ \cite{bochkarev1995scheme} and determined in the PMC configuration in ref.~\cite{yu2021new}. On the other hand, the difference between the top PMC masses for the two initial schemes has a value of $\Delta M_t^{\text{PMC}}=9.7$ GeV.

\begin{table*}[htbp]
\begin{tabular}{ccccccccccc}
\hline
 &$\quad$ & $H_{t}^{\left( 1\right) }$ & $H_{t}^{\left( 2\right) }$ & $H_{t}^{\left(
3\right) }$ & $H_{t}^{\left( 4\right) }$ & $\quad$ &$\overline{H}_{t}^{\left( 1\right) }$ & $\overline{H}_{t}^{\left( 2\right) }$ & $\overline{H}_{t}^{\left(
3\right) }$ & $\overline{H}_{t}^{\left( 4\right) }$  \\ \hline
Conv. & & $1.04595$ & $1.05574$ & $1.05876$ & $1.05995$
& & $0.949889$ & $0.93699$ & $0.932718$ & $0.930977$ \\
PMCs & & $1.04788$ & $1.04441$ & $1.04911$ & $1.04941$ &  & $0.948011$ & $0.950753$ & $0.944502$ & $0.944141$\\
 PMC & &  $1.04952$ & $1.04456 $ & $1.04999$ & $1.05035$&  &$0.94571$ & $0.950193$ & $0.94278$ & $0.942327$ \\ \hline
\end{tabular}
\caption{Numerical results for the top quark $H_t$ mass ratio and the inverse $\overline{H}_t$ ratio, under conventional (Conv.) scale setting, PMCs and PMCs corrected up to four loops, respectively. $\mu_r^{init}=80$ GeV (and $\mu_r^{init}=M_z$).}
\label{tableT}
\end{table*}

\begin{table}[H]
\centering
\begin{tabular}{ccccccc}
\hline
 & & $M_t$ (GeV) & & $\overline{m}_t$ (GeV)& & $\left\vert \Delta M_t \right \vert $ (GeV)\\ \hline
Conv.& &$173.8^{+1.3}_{-1.0}$ & & $160.6^{+1,8}_{-1.4}$ & & $13.2097$ \\
PMCs & &$172.10\pm 0.06$ & & $162.86^{+0.08}_{-0.09}$ & & $9.23861$ \\
PMC & &$172.3\pm 0.6$ & & $162.6\pm 0.7$ & &$9.70624$\\ \hline
\end{tabular}
\caption{The values obtained for the pole mass with input Eq.\eqref{inputmsbar2}, the running mass with input Eq.\eqref{inputpole}, and the difference between them for the top quark using the conventional setting (Conv.), PMCs, and PMC.}
\label{t-mass}
\end{table}

\section{Application: The PMC approach for electroweak parameter $\rho$} 
\label{sec:applications}

The parameter $\rho$ is used to describe the relationship between the charged and neutral currents in the weak interaction \cite{Veltman:1977kh}. In particular, $\rho$ is related to the ratio of the masses of the $W$ and $Z$ bosons and is used to predict changes in the $W$ boson mass and the effective weak leptonic mixing angle due to loop corrections. It can be written as
\begin{equation}
\rho=1+\delta \rho,
\end{equation}
with 
\begin{equation}
\delta \rho = \frac{\Pi_{ZZ}(0)}{M_Z^2} - \frac{\Pi_{WW}(0)}{M_W^2},
\end{equation}
where $\Pi_{ZZ}(0)$ and $\Pi_{WW}(0)$ are the transverse fractions of the $W$ and $Z$ boson self-energies evaluated for vanishing external momentum.

The precision in determining $\rho$ is crucial for verifying the internal consistency of the Standard Model and for searching for new physics beyond it. The $\delta \rho$ parameter for the on-shell definition of the top quark mass can be expressed as

\begin{equation}
\delta \rho_{\text{OS}}= \rho_0 \delta P(\mu),
\end{equation}
with the factor $\rho_0$ defined as, 
\begin{equation}
 \rho_0 = \frac{N_C G_F}{8\sqrt{2}\pi^2}M_t^2.
\label{eq:rho0}
\end{equation} 
The Fermi constant is given by \hbox{$G_F = 1.16638\times 10^{-5}\ \text{GeV}^{-2}$} \cite{particle2022review} and $N_C$ is the color factor with $N_C = 3$ for quarks and $N_C = 1$ for leptons. Meanwhile, the QCD corrections up to four-loop order are given by $\delta P$ \cite{Halzen:1991ik,Chetyrkin:1995ix,Schroder:2005db,Boughezal:2006xk,Chetyrkin:2006bj,Faisst:2006sr}:
\begin{equation}
\delta P \paren{\mu}= 1+d_1 a_s(\mu)+d_2 a_s^2(\mu)+d_3 a_s^3(\mu),
\end{equation}
where the coefficients $d_i$ are included in the Appendix \ref{app:coefficients}.


The goal of this section is to use the PMC scale setting procedure to fine-tune the values of $\rho_0$ and $\delta P$\footnote{In the reference \cite{yu2021new}, a PMC investigation of the factor $\delta P$ was performed in the single-scale case. In the present work, we propose to use the PMC approach to adjust both $\rho_0$ and $\delta P$ in both multi-scale and single-scale settings.} The determination of the top quark mass is usually crucial for the calculation of $\delta \rho$. In our study, the determination of $M_t(\mu)$ involves optimizing the scale setting of the observable $H_q$ in Eq.\eqref{Hrel}, where the PMC fit yields $M_t$. Using the calculations presented in subsection \ref{subsection:top}, Eq.\eqref{eq:rho0} can be expressed as follows:

\begin{equation} \label{rho0}
\begin{array}{ll}
\rho_0 = 9.282\pm 0.009\times 10^{-3},& \text{(PMCs)}\\
\rho_0 = 9.30\pm 0.09\times 10^{-3},& \text{(PMC)}\\
\rho_0 = 9.47_{-0.14}^{+0.11}\times 10^{-3}.& \text{(Conventional)}
\end{array}
\end{equation}

On the other hand, the PMC approach for $\delta P$ gives\begin{eqnarray}
\delta P^{\text{PMC}} &= & 1+r_1^{\text{conf}} \alpha_s\paren{Q_1}+r_2^{\text{conf}} \alpha_s^2\paren{Q_2}\\
\nonumber && +r_3^{\text{conf}} \alpha_s^3\paren{Q_3}.
\end{eqnarray}
Here $r_i^{\text{conf}}$ represents the conformal coefficients given in the appendix \ref{app:coefficients}. These conformal coefficients are independent of the choice of the initial renormalization scale. Using eqs.\eqref{multiscale}-\eqref{rrr3}, in the multiscale PMC setting we obtain: $Q_1=26.11\ \text{GeV}$, $Q_2=84.08\ \text{GeV}$, and $Q_3$ is fixed as the last determined scale.

In the single-scale PMCs setting we have 
\begin{eqnarray}
\delta P^{\text{PMCs}} &= & 1+r_1^{\text{conf}} \alpha_s\paren{Q_s^{\text{NLL}}}+r_2^{\text{conf}} \alpha_s^2\paren{Q_s^{\text{NLL}}}\\
\nonumber && +r_3^{\text{conf}} \alpha_s^3\paren{Q_s^{\text{NLL}}},
\end{eqnarray}
where $Q_s^{\text{LL}}=26.43\ \text{GeV}$ and $Q_s^{\text{NLL}}=23.96\ \text{GeV}$ (see Appendix \ref{single}).

\begin{figure}[]
\centering
\includegraphics[scale=.57]{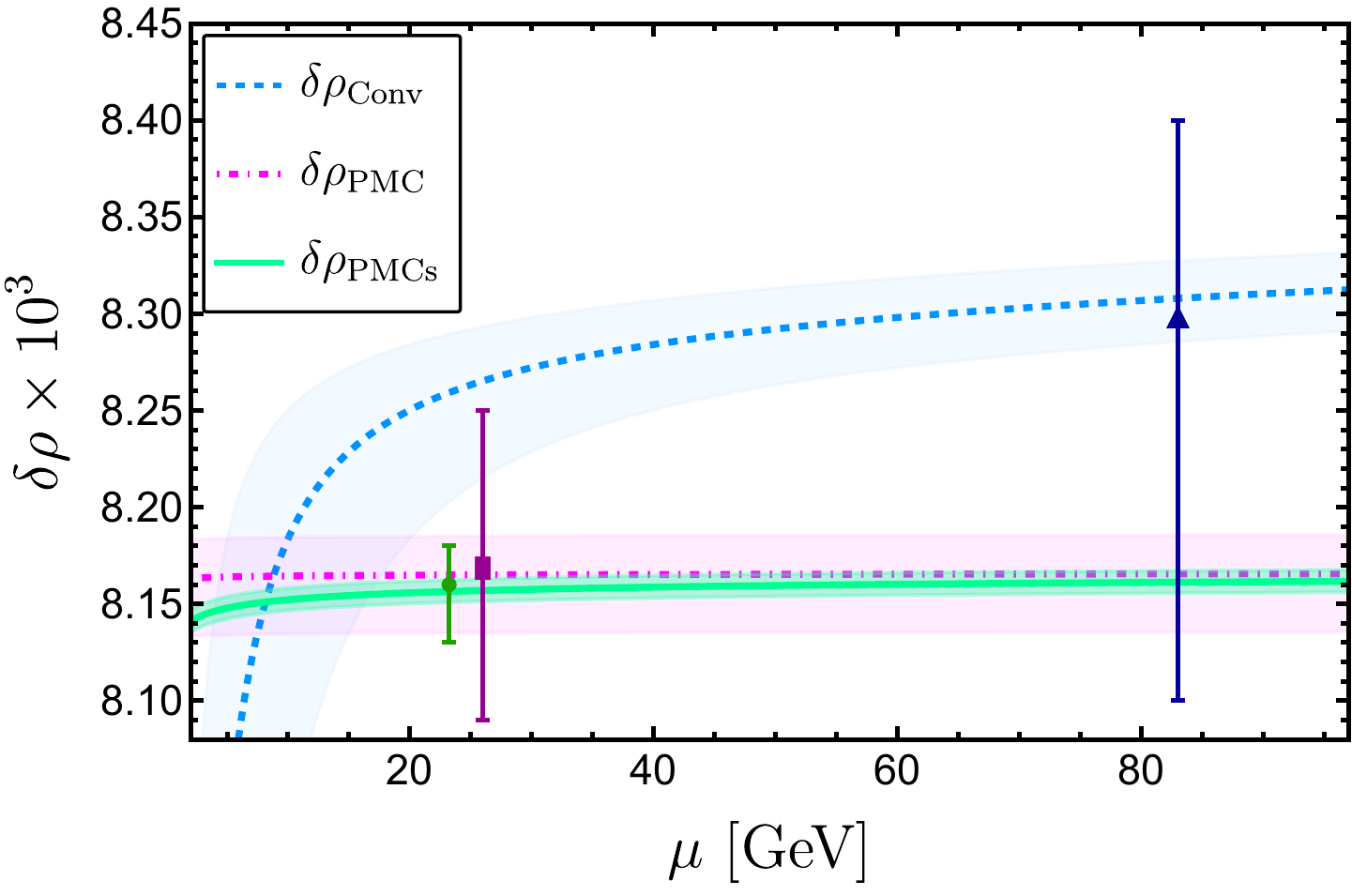}
\caption{Renormalization scale dependence of $\delta \rho$. Dashed line for conventional scale setting, dotted line for multi-scale PMC setting, and solid line for single-scale PMCs setting.}
\label{plotRHO}
\end{figure}

Numerically, the renormalization scale dependence for the QCD four-loop correction of the electroweak parameter $\delta\rho$ is shown in Fig.~\ref{plotRHO}. The dashed line represents a higher renormalization scale dependence, corresponding to the conventional scaling calculation. The dotted and solid lines represent the multi- and single-scale PMC settings. In all cases, the shaded region corresponds to the uncertainty of the calculations.

Due to the perturbative nature of the single-scale approach, we observe that a minimal initial residual dependence is stabilized at higher energies compared to the calculation performed with the multi-scale approach. The significant stability of $\delta \rho$ achieved with the PMC approach is attributed to the elimination of ambiguity in the choice of the renormalization scale and the absorption of nonconformal terms into the effective running coupling.

The triangle, box, and circle points of the plot correspond to the central values of $\delta \rho$ obtained from the conventional scale setting, multi-scale PMC, and single-scale PMCs, respectively. The same input parameters used in this section and in section \ref{sec:num} have been used to obtain the following results:
\begin{equation} \label{rho0}
\begin{array}{ll}
\delta \rho  =  8.16_{-0.03}^{+0.02} \times 10^{-3},& \text{(PMCs)}\\
\delta \rho = 8.17 \pm 0.08\times 10^{-3},& \text{(PMC)}\\
\delta \rho = 8.3^{+0.1}_{-0.2} \times 10^{-3}.& \text{(Conventional)}
\end{array}
\end{equation}
The error in the conventional case is determined by taking $\mu = [M_t/2, 2M_t]$ as the range of renormalization scales, and in the PMC case, it is obtained by considering the determination of the single scale at NLL and LL orders, as well as other error propagations.


\section{Summary}
\label{sec:summary}
The conventional determination of the mass ratio for heavy quarks assigns an arbitrary range and systematic error to their pQCD corrections at fixed order, breaking the renormalization group invariance. The PMC formalism provides a solution to the aforementioned problem with respect to the relation between the pole and the running masses by systematically adjusting the renormalization scale of the process. The results of the PMC formalism are independent of the initial renormalization scale. The PMC scale of the process is determined from the absorption process of the nonconformal terms, which is a significant theoretical advantage since they are determined and not arbitrarily chosen. A second improvement is that at each order, a different scale adjustment is presented, which helps the series to converge; in cases where the series converges slowly, it is helpful to introduce a single universal PMC scale, which improves the determination of the observable. In this work, this situation is present in the determination of the mass ratio for the charm quark, where the prediction made with the single scale adjustment is much better than in the other case.

The results obtained in this work are \hbox{$M_b^{\text{PMC}}=4.86_{-0.02}^{+0.03}\ \text{GeV}$}, $M_t^{\text{PMC}}=172.3 \pm 0.6\ \text{GeV}$ and $\overline{m}_t^{\text{PMC}}=162.6\pm 0. 7$ GeV, improving the accuracy that can be obtained in some tests of the Standard Model and the sensitivity to new phenomena, since the formalism eliminates the ambiguities of the scheme and the renormalization scale. The pole mass PMCs can be used in the threshold phenomena \cite{penin1999top,hoang2000top}, and a better correction can influence the predictions for the cross section of quark production.

Furthermore, as an application, in this work we focus on the importance of the $\rho$ parameter in the description of the weak interactions and its relation to the masses of the $W$ and $Z$ bosons. Accuracy in the determination of $\rho$ is crucial for verifying the internal consistency of the Standard Model and for searching for new physics. To achieve this, the PMC scaling procedure performed for the top quark mass in Sec. \ref{sec:num} is used to tune the values of $\rho_0$ and $\delta P$.
In particular, it is found that the determination of the parameter $\delta \rho$ is between $8.16_{-0.03}^{+0.02}$ and $8.3^{+0.1}_{-0.2} \times 10^{-3}$ in the PMC approach, compared to $8.3^{+0.1}_{-0.2} \times 10^{-3}$ for the conventional approach. These results highlight the improved stability and accuracy achieved by using PMC, as seen in Fig.~\ref{plotRHO}. Overall, this section highlights the importance of PMC scale fitting in determining crucial parameters in the Standard Model and its ability to eliminate ambiguities in the choice of renormalization scale, thereby improving the accuracy of theoretical predictions. The results obtained have significant implications for tests of the Standard Model and the search for new physics in future experiments.
\vspace{0.3cm}

\textbf{Acknowledgements:} This work was supported by Project PIIC No. 026/2019 DPP-USM, and by ANID PIA/APOYO AFB180002 (Chile), and by partial support from FONDECYT (Chile) under Grant No. 1230391. We would like to thank C. Dib for revising this manuscript, G. Cvetič for his contribution to improving this article, and also J. Echeverria for useful comments and discussions.

\emph{It is with great sadness that I bid farewell to Professor Iván Schmidt Andrade, who passed away during the course of this work. His discipline, perseverance and way of looking at problems led him to mark a before and after in physics. Farewell, Professor, wherever you are, and thank you for your patience and training.}

\appendix 

\section{\label{single}Coefficients for the PMC single-scale setting up to NNLO}
The single scale renormalization PMC is
\begin{equation}
\ln \left(\frac{Q_{s}}{\mu_r^{init}}\right)^2=T_0+T_1 a_s(\mu_r^{init})+T_2 a_s^2(\mu_r^{init})+\mathcal{O}\paren{a_s^3},
\end{equation}
where the $T_i$ are process-dependent coefficients. We can get $Q_s$ at NNLL order, whose three coefficients are

\begin{widetext}
\begin{equation}
\begin{aligned}
T_0=&-\frac{r_{2,1}}{r_{1,0}}, \\
T_1=& \frac{(p+1)\left(r_{2,0} r_{2,1}-r_{1,0} r_{3,1}\right)}{p\ r_{1,0}^2}+\frac{(p+1)\left(r_{2,1}^2-r_{1,0} r_{3,2}\right)}{2 r_{1,0}^2} \beta_0, \\
T_2=& \frac{(p+1)^2\left(r_{1,0} r_{2,0} r_{3,1}-r_{2,0}^2 r_{2,1}\right)+p(p+2)\left(r_{1,0} r_{2,1} r_{3,0}-r_{1,0}^2 r_{4,1}\right)}{p^2 r_{1,0}^3}+\frac{(p+2)\left(r_{2,1}^2-r_{1,0} r_{3,2}\right)}{2 r_{1,0}^2} \beta_1 \\
&-\frac{p(p+1) r_{2,0} r_{2,1}^2+(p+1)^2\left(r_{2,0} r_{2,1}^2-2 r_{1,0} r_{2,1} r_{3,1}-r_{1,0} r_{2,0} r_{3,2}\right)+(p+1)(p+2) r_{1,0}^2 r_{4,2}}{2 p\  r_{1,0}^3} \beta_0 \\
&+\frac{(p+1)(p+2)\left(r_{1,0} r_{2,1} r_{3,2}-r_{1,0}^2 r_{4,3}\right)+(p+1)(1+2 p)\left(r_{1,0} r_{2,1} r_{3,2}-r_{2,1}^3\right)}{6 r_{1,0}^3} \beta_0^2 .
\end{aligned}
\end{equation}
\end{widetext}

\section{Complete and PMC QCD correction coefficients at the four-loop level of the  $\delta P$}
\label{app:coefficients}

Here, we first include the perturbative coefficients of the observable $\delta P$ introduced in Sec. \ref{sec:applications}.

\begin{widetext} 
\begin{eqnarray}
d_1 & = & -2.14 C_F,\\
d_2 & = & -4.42 C_F + 7.23 C_F^2  + C_A C_F \left(-6.29 - 31.5 \ln \frac{\mu^2}{M_t^2} \right) + N_f C_F \left( 21.4 + 0.36 \ln \frac{\mu^2}{M_t^2} \right),\\
d_3 & = & -0.785 C_F^3 + C_F^2 T \left( 8.30 +   N_f \left( -8.74 - 1.62\ln \frac{\mu^2}{M_t^2}\right)\right)+ C_A C_F^2 \left( 17.2 + 5.92 \ln \frac{\mu^2}{M_t^2} \right)\\
\nonumber & &+  C_A^2 C_F \left(-31.0 - \ln \frac{\mu^2}{M_t^2} - 1.80 \ln^2 \frac{\mu^2}{M_t^2} \right) +C_F T^2 \left( -3.86 + N_f \left(7.61 - 1.69 \ln \frac{\mu^2}{M_t^2} \right. \right) \\
\nonumber & & -  6.4\times 10^{-10} \ln \frac{\mu^2}{M_t^2} + 10^{-10} \ln^2 \frac{\mu^2}{M_t^2} + N_f^2 \left( -3.45 + 1.79 \ln \frac{\mu^2}{M_t^2} -0.24 \ln^2 \frac{\mu^2}{M_t^2} \right) \\
\nonumber & & + C_A C_F T \left(-25.4 - 4.64 \ln \frac{\mu^2}{M_t^2} + N_f \left(24.8 + \ln \frac{\mu^2}{M_t^2} +1.31 \ln^2 \frac{\mu^2}{M_t^2} \right)\right).
\end{eqnarray}
The PMC conformal coefficients $r_i^{\text{conf}}$ for the pertubative serie of $\delta P$ up to four-loop level, is given by
\begin{eqnarray}
r_1^{\text{conf}} &=& -0.68 C_F ,\\
r_2^{\text{conf}} &=& -0.45 C_F +0.73 C_F^2-0.64C_A C_F +\frac{5.96 C_A C_F}{T},\\
r_3^{\text{conf}} &=&\frac{C_F}{T} \left(C_A^2 (-29.6
   T-3.6)+C_A C_F (16.4   T-5.69)-24.7 C_A T^2+T
   \left(-0.78 C_F^2+8.30 C_F T-3.86   T^2\right)\right),
\end{eqnarray}
\end{widetext}

where $C_F=4/3 $, $C_A=3 $, and $T=1/2 $ are the known $SU(3)$ color factors.

\bibliographystyle{utphys}
\bibliography{refquarkmass.bib}

\providecommand{\href}[2]{#2}\begingroup\raggedright\begin{thebibliography}{10}

\bibitem{gray1990three}
N.~Gray, D.~J. Broadhurst, W.~Grafe, and K.~Schilcher, ``{Three Loop Relation
  of Quark (Modified) Ms and Pole Masses},''
  \href{http://dx.doi.org/10.1007/BF01614703}{{\em Z. Phys. C} {\bfseries 48}
  (1990) 673--680}.

\bibitem{melnikov2000three}
K.~Melnikov and T.~v. Ritbergen, ``{The Three loop relation between the MS-bar
  and the pole quark masses},''
  \href{http://dx.doi.org/10.1016/S0370-2693(00)00507-4}{{\em Phys. Lett. B}
  {\bfseries 482} (2000) 99--108},
  \href{http://arxiv.org/abs/hep-ph/9912391}{{\ttfamily arXiv:hep-ph/9912391}}.

\bibitem{marquard2015quark}
P.~Marquard, A.~V. Smirnov, V.~A. Smirnov, and M.~Steinhauser, ``{Quark Mass
  Relations to Four-Loop Order in Perturbative QCD},''
  \href{http://dx.doi.org/10.1103/PhysRevLett.114.142002}{{\em Phys. Rev.
  Lett.} {\bfseries 114} no.~14, (2015) 142002},
  \href{http://arxiv.org/abs/1502.01030}{{\ttfamily arXiv:1502.01030
  [hep-ph]}}.

\bibitem{Steinhauser2016xkf}
M.~Steinhauser, \href{http://dx.doi.org/10.22323/1.265.0148}{``{Relation
  between the pole and $\overline {MS}$ quark mass in QCD},''} in {\em PoS
  DIS2016}, p.~148.
\newblock 2016.

\bibitem{kataev2020multiloop}
A.~L. Kataev and V.~S. Molokoedov, ``{Multiloop contributions to the on-shell-$
  \overline{{\mathrm{MS}}}$ heavy quark mass relation in QCD and the asymptotic
  structure of the corresponding series: the updated consideration},''
  \href{http://dx.doi.org/10.1140/epjc/s10052-020-08673-6}{{\em Eur. Phys. J.
  C} {\bfseries 80} no.~12, (2020) 1160},
  \href{http://arxiv.org/abs/1807.05406}{{\ttfamily arXiv:1807.05406
  [hep-ph]}}.

\bibitem{brodsky2012scale}
S.~J. Brodsky and X.~G. Wu, ``{Scale Setting Using the Extended Renormalization
  Group and the Principle of Maximum Conformality: the QCD Coupling Constant at
  Four Loops},'' \href{http://dx.doi.org/10.1103/PhysRevD.85.034038}{{\em Phys.
  Rev. D} {\bfseries 85} (2012) 034038},
  \href{http://arxiv.org/abs/1111.6175}{{\ttfamily arXiv:1111.6175 [hep-ph]}}.
  [erratum: Phys. Rev. D \textbf{86}, 079903 (2012)].

\bibitem{brodsky2012setting}
S.~J. Brodsky and L.~Di~Giustino, ``{Setting the Renormalization Scale in QCD:
  The Principle of Maximum Conformality},''
  \href{http://dx.doi.org/10.1103/PhysRevD.86.085026}{{\em Phys. Rev. D}
  {\bfseries 86} (2012) 085026},
  \href{http://arxiv.org/abs/1107.0338}{{\ttfamily arXiv:1107.0338 [hep-ph]}}.

\bibitem{brodsky2012eliminating}
S.~J. Brodsky and X.~G. Wu, ``{Eliminating the Renormalization Scale Ambiguity
  for Top-Pair Production Using the Principle of Maximum Conformality},''
  \href{http://dx.doi.org/10.1103/PhysRevLett.109.042002}{{\em Phys. Rev.
  Lett.} {\bfseries 109} (2012) 042002},
  \href{http://arxiv.org/abs/1203.5312}{{\ttfamily arXiv:1203.5312 [hep-ph]}}.

\bibitem{mojaza2013systematic}
M.~Mojaza, S.~J. Brodsky, and X.-G. Wu, ``{Systematic All-Orders Method to
  Eliminate Renormalization-Scale and Scheme Ambiguities in Perturbative
  QCD},'' \href{http://dx.doi.org/10.1103/PhysRevLett.110.192001}{{\em Phys.
  Rev. Lett.} {\bfseries 110} (2013) 192001},
  \href{http://arxiv.org/abs/1212.0049}{{\ttfamily arXiv:1212.0049 [hep-ph]}}.

\bibitem{brodsky2014systematic}
S.~J. Brodsky, M.~Mojaza, and X.-G. Wu, ``{Systematic Scale-Setting to All
  Orders: The Principle of Maximum Conformality and Commensurate Scale
  Relations},'' \href{http://dx.doi.org/10.1103/PhysRevD.89.014027}{{\em Phys.
  Rev. D} {\bfseries 89} (2014) 014027},
  \href{http://arxiv.org/abs/1304.4631}{{\ttfamily arXiv:1304.4631 [hep-ph]}}.

\bibitem{wu2013renormalization}
X.-G. Wu, S.~J. Brodsky, and M.~Mojaza, ``{The Renormalization Scale-Setting
  Problem in QCD},'' \href{http://dx.doi.org/10.1016/j.ppnp.2013.06.001}{{\em
  Prog. Part. Nucl. Phys.} {\bfseries 72} (2013) 44--98},
  \href{http://arxiv.org/abs/1302.0599}{{\ttfamily arXiv:1302.0599 [hep-ph]}}.

\bibitem{singlescale1}
J.-M. Shen, X.-G. Wu, B.-L. Du, and S.~J. Brodsky, ``{Novel All-Orders
  Single-Scale Approach to QCD Renormalization Scale-Setting},''
  \href{http://dx.doi.org/10.1103/PhysRevD.95.094006}{{\em Phys. Rev. D}
  {\bfseries 95} no.~9, (2017) 094006},
  \href{http://arxiv.org/abs/1701.08245}{{\ttfamily arXiv:1701.08245
  [hep-ph]}}.

\bibitem{brodsky1983elimination}
S.~J. Brodsky, G.~P. Lepage, and P.~B. Mackenzie, ``{On the Elimination of
  Scale Ambiguities in Perturbative Quantum Chromodynamics},''
  \href{http://dx.doi.org/10.1103/PhysRevD.28.228}{{\em Phys. Rev. D}
  {\bfseries 28} (1983) 228}.

\bibitem{bollini1972dimensional}
C.~G. Bollini and J.~J. Giambiagi, ``{Dimensional Renormalization: The Number
  of Dimensions as a Regularizing Parameter},''
  \href{http://dx.doi.org/10.1007/BF02895558}{{\em Nuovo Cim. B} {\bfseries 12}
  (1972) 20--26}.

\bibitem{veltman1972regularization}
G.~'t~Hooft and M.~J.~G. Veltman, ``{Regularization and Renormalization of
  Gauge Fields},'' \href{http://dx.doi.org/10.1016/0550-3213(72)90279-9}{{\em
  Nucl. Phys. B} {\bfseries 44} (1972) 189--213}.

\bibitem{brodsky2012self}
S.~J. Brodsky and X.-G. Wu, ``{Self-Consistency Requirements of the
  Renormalization Group for Setting the Renormalization Scale},''
  \href{http://dx.doi.org/10.1103/PhysRevD.86.054018}{{\em Phys. Rev. D}
  {\bfseries 86} no.~5, (2012) 054018},
  \href{http://arxiv.org/abs/1208.0700}{{\ttfamily arXiv:1208.0700 [hep-ph]}}.

\bibitem{CSR}
H.~J. Lu and S.~J. Brodsky, ``{Commensurate scale relations in quantum
  chromodynamics},'' \href{http://dx.doi.org/10.1016/0920-5632(95)00093-O}{{\em
  Nucl. Phys. B Proc. Suppl.} {\bfseries 39BC} (1995) 309--311}.

\bibitem{marquard2016ms}
P.~Marquard, A.~V. Smirnov, V.~A. Smirnov, M.~Steinhauser, and D.~Wellmann,
  ``{$\overline{\rm MS}$-on-shell quark mass relation up to four loops in QCD
  and a general SU$(N)$ gauge group},''
  \href{http://dx.doi.org/10.1103/PhysRevD.94.074025}{{\em Phys. Rev. D}
  {\bfseries 94} no.~7, (2016) 074025},
  \href{http://arxiv.org/abs/1606.06754}{{\ttfamily arXiv:1606.06754
  [hep-ph]}}.

\bibitem{Kataev:2016jai}
A.~L. Kataev and V.~S. Molokoedov, ``{From perturbative calculations of the QCD
  static potential towards four-loop pole-running heavy quarks masses
  relation},'' \href{http://dx.doi.org/10.1088/1742-6596/762/1/012078}{{\em J.
  Phys. Conf. Ser.} {\bfseries 762} no.~1, (2016) 012078},
  \href{http://arxiv.org/abs/1604.03485}{{\ttfamily arXiv:1604.03485
  [hep-ph]}}.

\bibitem{Kataev:2015gvt}
A.~L. Kataev and V.~S. Molokoedov, ``{On the flavour dependence of the
  $\mathcal{O}(\alpha_s^4)$ correction to the relation between running and pole
  heavy quark masses},''
  \href{http://dx.doi.org/10.1140/epjp/i2016-16271-7}{{\em Eur. Phys. J. Plus}
  {\bfseries 131} no.~8, (2016) 271},
  \href{http://arxiv.org/abs/1511.06898}{{\ttfamily arXiv:1511.06898
  [hep-ph]}}.

\bibitem{baikov2017five}
P.~A. Baikov, K.~G. Chetyrkin, and J.~H. K\"uhn, ``{Five-Loop Running of the
  QCD coupling constant},''
  \href{http://dx.doi.org/10.1103/PhysRevLett.118.082002}{{\em Phys. Rev.
  Lett.} {\bfseries 118} no.~8, (2017) 082002},
  \href{http://arxiv.org/abs/1606.08659}{{\ttfamily arXiv:1606.08659
  [hep-ph]}}.

\bibitem{particle2022review}
R.~L. P. D.~G. Workman, ``{Review of Particle Physics},''
  \href{http://dx.doi.org/10.1093/ptep/ptac097}{{\em PTEP} {\bfseries 2022}
  (2022) 083C01}.

\bibitem{hoang2006charm}
A.~H. Hoang and A.~V. Manohar, ``{Charm quark mass from inclusive semileptonic
  B decays},'' \href{http://dx.doi.org/10.1016/j.physletb.2005.12.020}{{\em
  Phys. Lett. B} {\bfseries 633} (2006) 526--532},
  \href{http://arxiv.org/abs/hep-ph/0509195}{{\ttfamily arXiv:hep-ph/0509195
  [hep-ph]}}.

\bibitem{chetyrkin2000relation}
K.~G. Chetyrkin and M.~Steinhauser, ``{The Relation between the
  $\overline{\text{MS}}$ and the on-shell quark mass at order $\alpha_s^3$},''
  \href{http://dx.doi.org/10.1016/S0550-3213(99)00784-1}{{\em Nucl. Phys. B}
  {\bfseries 573} (2000) 617--651},
  \href{http://arxiv.org/abs/hep-ph/9911434}{{\ttfamily arXiv:hep-ph/9911434
  [hep-ph]}}.

\bibitem{hoang2020top}
A.~H. Hoang, ``{What is the Top Quark Mass?},''
  \href{http://dx.doi.org/10.1146/annurev-nucl-101918-023530}{{\em Ann. Rev.
  Nucl. Part. Sci.} {\bfseries 70} (2020) 225--255},
  \href{http://arxiv.org/abs/2004.12915}{{\ttfamily arXiv:2004.12915
  [hep-ph]}}.

\bibitem{jegerlehner2003alpha}
F.~Jegerlehner and M.~Y. Kalmykov, ``{$\mathcal{O}\left(\alpha \alpha_s
  \right)$ relation between pole- and $\overline{\text{MS}}$- mass of the $t$
  quark},'' {\em Acta Phys. Polon. B} {\bfseries 34} (2003) 5335--5344,
  \href{http://arxiv.org/abs/hep-ph/0310361}{{\ttfamily arXiv:hep-ph/0310361
  [hep-ph]}}.

\bibitem{kataev2022notes}
A.~L. Kataev and V.~S. Molokoedov, ``{Notes on Interplay between the QCD and EW
  Perturbative Corrections to the Pole-Running-to-Top-Quark Mass Ratio},''
  \href{http://dx.doi.org/10.1134/S0021364022600902}{{\em JETP Lett.}
  {\bfseries 115} no.~12, (2022) 704--712},
  \href{http://arxiv.org/abs/2201.12073}{{\ttfamily arXiv:2201.12073
  [hep-ph]}}.

\bibitem{wang2018precise}
S.~Q. Wang, X.~G. Wu, Z.~G. Si, and S.~J. Brodsky, ``{A precise determination
  of the top-quark pole mass},''
  \href{http://dx.doi.org/10.1140/epjc/s10052-018-5688-1}{{\em Eur. Phys. J. C}
  {\bfseries 78} no.~3, (2018) 237},
  \href{http://arxiv.org/abs/1703.03583}{{\ttfamily arXiv:1703.03583
  [hep-ph]}}.

\bibitem{bochkarev1995scheme}
A.~I. Bochkarev and R.~S. Willey, ``{On the scheme dependence of the
  electroweak radiative corrections},''
  \href{http://dx.doi.org/10.1103/PhysRevD.51.R2049}{{\em Phys. Rev. D}
  {\bfseries 51} (1995) 2049--2052},
  \href{http://arxiv.org/abs/hep-ph/9407261}{{\ttfamily arXiv:hep-ph/9407261
  [hep-ph]}}.

\bibitem{yu2021new}
Q.~Yu, H.~Zhou, J.~Yan, X.~D. Huang, and X.~G. Wu, ``{A new analysis of the
  pQCD contributions to the electroweak parameter \ensuremath{\rho} using the
  single-scale approach of principle of maximum conformality},''
  \href{http://dx.doi.org/10.1016/j.physletb.2021.136574}{{\em Phys. Lett. B}
  {\bfseries 820} (2021) 136574},
  \href{http://arxiv.org/abs/2105.07230}{{\ttfamily arXiv:2105.07230
  [hep-ph]}}.

\bibitem{Veltman:1977kh}
M.~J.~G. Veltman, ``{Limit on Mass Differences in the Weinberg Model},''
  \href{http://dx.doi.org/10.1016/0550-3213(77)90342-X}{{\em Nucl. Phys. B}
  {\bfseries 123} (1977) 89--99}.

\bibitem{Halzen:1991ik}
F.~Halzen, B.~A. Kniehl, and M.~L. Stong, ``{Two loop electroweak
  parameters},'' \href{http://dx.doi.org/10.1007/BF01554085}{{\em Z. Phys. C}
  {\bfseries 58} (1993) 119--132}.

\bibitem{Chetyrkin:1995ix}
K.~G. Chetyrkin, J.~H. Kuhn, and M.~Steinhauser, ``{Corrections of order ${\cal
  O}(G_F M_t^2 \alpha_s^2)$ to the $\rho$ parameter},''
  \href{http://dx.doi.org/10.1016/0370-2693(95)00380-4}{{\em Phys. Lett. B}
  {\bfseries 351} (1995) 331--338},
  \href{http://arxiv.org/abs/hep-ph/9502291}{{\ttfamily arXiv:hep-ph/9502291}}.

\bibitem{Schroder:2005db}
Y.~Schroder and M.~Steinhauser, ``{Four-loop singlet contribution to the rho
  parameter},'' \href{http://dx.doi.org/10.1016/j.physletb.2005.06.085}{{\em
  Phys. Lett. B} {\bfseries 622} (2005) 124--130},
  \href{http://arxiv.org/abs/hep-ph/0504055}{{\ttfamily arXiv:hep-ph/0504055}}.

\bibitem{Boughezal:2006xk}
R.~Boughezal and M.~Czakon, ``{Single scale tadpoles and ${\cal O}(G_F M_t^2
  \alpha_s^2)$ corrections to the $\rho$ parameter},''
  \href{http://dx.doi.org/10.1016/j.nuclphysb.2006.08.007}{{\em Nucl. Phys. B}
  {\bfseries 755} (2006) 221--238},
  \href{http://arxiv.org/abs/hep-ph/0606232}{{\ttfamily arXiv:hep-ph/0606232}}.

\bibitem{Chetyrkin:2006bj}
K.~G. Chetyrkin, M.~Faisst, J.~H. Kuhn, P.~Maierhofer, and C.~Sturm,
  ``{Four-Loop QCD Corrections to the Rho Parameter},''
  \href{http://dx.doi.org/10.1103/PhysRevLett.97.102003}{{\em Phys. Rev. Lett.}
  {\bfseries 97} (2006) 102003},
  \href{http://arxiv.org/abs/hep-ph/0605201}{{\ttfamily arXiv:hep-ph/0605201}}.

\bibitem{Faisst:2006sr}
M.~Faisst, P.~Maierhoefer, and C.~Sturm, ``{Standard and epsilon-finite Master
  Integrals for the rho-Parameter},''
  \href{http://dx.doi.org/10.1016/j.nuclphysb.2006.12.014}{{\em Nucl. Phys. B}
  {\bfseries 766} (2007) 246--268},
  \href{http://arxiv.org/abs/hep-ph/0611244}{{\ttfamily arXiv:hep-ph/0611244}}.

\bibitem{penin1999top}
A.~A. Penin and A.~A. Pivovarov, ``{Top quark threshold production in gamma
  gamma collision in the next-to-leading order},''
  \href{http://dx.doi.org/10.1016/S0550-3213(99)00205-9}{{\em Nucl. Phys. B}
  {\bfseries 550} (1999) 375--396},
  \href{http://arxiv.org/abs/hep-ph/9810496}{{\ttfamily arXiv:hep-ph/9810496
  [hep-ph]}}.

\bibitem{hoang2000top}
A.~H. Hoang, M.~Beneke, K.~Melnikov, T.~Nagano, A.~Ota, A.~A. Penin, A.~A.
  Pivovarov, A.~Signer, V.~A. Smirnov, and Y.~Sumino, ``{Top - anti-top pair
  production close to threshold: Synopsis of recent NNLO results},''
  \href{http://dx.doi.org/10.1007/s1010500c0003}{{\em Eur. Phys. J. direct}
  {\bfseries 2} no.~1, (2000) 3},
  \href{http://arxiv.org/abs/hep-ph/0001286}{{\ttfamily arXiv:hep-ph/0001286
  [hep-ph]}}.

\end{thebibliography}\endgroup

\end{document}